\documentclass[acmlarge]{acmart}
\usepackage[utf8]{inputenc}

\AtBeginDocument{%
  }

\setcopyright{cc}
\setcctype{by}
\copyrightyear{2026}
\acmYear{2026}
\acmDOI{10.1145/3857988}

\acmJournal{IMWUT}
\acmVolume{10}
\acmNumber{4}
\acmArticle{219}
\acmMonth{12}

\usepackage{longtable, booktabs, array}

\usepackage{graphicx}
\usepackage{threeparttable}
\usepackage{float}
\usepackage{placeins}
\usepackage{setspace}
\usepackage{xcolor}
\usepackage{textgreek}

\usepackage{etoolbox}
\newtoggle{showrevisions}
\togglefalse{showrevisions}

\newcommand{\rev}[1]{%
  \iftoggle{showrevisions}{\textcolor{blue}{#1}}{#1}%
}

\begin{document}
\title{A Formative Study of Brief Affective Text as a Complement to Wearable Sensing for Longitudinal Student Health Monitoring}

\author{Tamunotonye Harry}
\email{tamunotonye.harry@uvm.edu}
\authornote{Corresponding author.}
\orcid{https://orcid.org/0009-0004-8854-7955}
\affiliation{%
  \institution{Department of Computer Science, University of Vermont}
  \city{Burlington}
  \state{Vermont}
  \country{USA}}

\author{Johanna Hidalgo}
\email{johanna.hidalgo@uvm.edu}
\orcid{https://orcid.org/0000-0002-2604-3741}
\affiliation{%
  \institution{Department of Psychological Science, University of Vermont}
  \city{Burlington}
  \state{Vermont}
  \country{USA}}

\author{Matthew Price}
\email{matthew.price@uvm.edu}
\orcid{https://orcid.org/0000-0001-5637-9230}
\affiliation{%
  \institution{Department of Psychological Science, University of Vermont}
  \city{Burlington}
  \state{Vermont}
  \country{USA}}

\author{Yuanyuan Feng}
\email{yuanyuan.feng@uvm.edu}
\orcid{https://orcid.org/0000-0002-6751-196X}
\affiliation{%
  \institution{Department of Computer Science, University of Vermont}
  \city{Burlington}
  \state{Vermont}
  \country{USA}}

\author{Kathryn Stanton}
\email{kathryn.stanton@uvm.edu}
\orcid{https://orcid.org/0009-0001-3822-7505}
\affiliation{%
  \institution{MassMutual Center of Excellence in Complex Systems and Data Science, University of Vermont}
  \city{Burlington}
  \state{Vermont}
  \country{USA}}

\author{Connie Tompkins}
\email{connie.tompkins@med.uvm.edu}
\orcid{https://orcid.org/0000-0002-6927-0792}
\affiliation{%
  \institution{Department of Rehabilitation and Movement Science, University of Vermont}
  \city{Burlington}
  \state{Vermont}
  \country{USA}}

\author{Peter Sheridan Dodds}
\email{peter.dodds@uvm.edu}
\orcid{https://orcid.org/0000-0003-1973-8614}
\affiliation{%
  \institution{MassMutual Center of Excellence in Complex Systems and Data Science, University of Vermont}
  \city{Burlington}
  \state{Vermont}
  \country{USA}}

\author{Mikaela Irene Fudolig}
\email{mikaela.fudolig@adelaide.edu.au}
\orcid{https://orcid.org/0000-0002-8960-9116}
\affiliation{%
  \institution{School of Mathematical Sciences, Adelaide University}
  \city{Adelaide}
  \country{Australia}}

\author{Laura Bloomfield}
\email{laura.bloomfield@gmail.com}
\orcid{https://orcid.org/0000-0003-0416-3440}
\affiliation{%
  \institution{Department of Mathematics and Statistics, University of Vermont}
  \city{Burlington}
  \state{Vermont}
  \country{USA}}

\author{Christopher Danforth}
\email{chris.danforth@uvm.edu}
\orcid{https://orcid.org/0000-0002-9857-2845}
\affiliation{%
  \institution{MassMutual Center of Excellence in Complex Systems and Data Science, University of Vermont}
  \city{Burlington}
  \state{Vermont}
  \country{USA}}


\begin{abstract}
Wearable devices capture physiological and behavioral data with increasing fidelity, but the psychological context shaping these outcomes is difficult to recover from sensor data alone, limiting the utility of passive sensing for digital health goals such as early detection of distress, personalized intervention, and timely clinical outreach. We examined whether ultra-brief naturalistic concern text could serve as a scalable complement to passive sensing. In a year-long study of 458 university students (3,610 person-waves) tracked with Oura rings, participants responded bimonthly to an open-ended prompt about what concerned them most; responses had a median length of three words. We compared dictionary-based, general pretrained, and domain-adapted NLP approaches using within-person mixed-effects models across nine sleep and physical activity outcomes to determine which method best recovers physiologically relevant signal from brief naturalistic text. Weeks dominated by academic concern framing were associated with lower physical activity; weeks characterized by emotional exhaustion language were associated with poorer sleep quality and lower heart rate variability. \rev{General pretrained embeddings performed as well as or better than domain-adapted models across most outcomes, with differences between the two generally small and within the range of estimation noise.} \rev{Zero-shot classification of concern topics showed no consistent evidence of association with outcomes; affective dimensions across all three methods showed more associations, though these did not survive correction for multiple comparisons, offering preliminary evidence that emotional register may carry more signal than topical content.} These findings offer design guidance: ultra-brief affective prompts enrich the psychological interpretability of passive physiological data at minimal burden.
\end{abstract}


\keywords{wearable sensing, passive sensing, natural language 
processing, ecological momentary assessment, digital health, 
affective computing, longitudinal studies, university students}



\settopmatter{printacmref=true}
\maketitle

\setlength{\headheight}{18.48402pt}
\section{Introduction}
\label{sec:intro}

It is Week 9 of the spring semester. A university student wakes up, syncs their Oura ring, and the data lands cleanly on a server: 6.8 hours of sleep, 82\% efficiency, heart rate variability during sleep of 54~ms, down from their personal average of 71~ms. Something is off. The wearable has detected it with precision. But it cannot tell you why. It cannot tell you whether the student is behind on a paper, grieving a friendship that ended, overwhelmed by financial aid paperwork, or simply fighting a cold. The ring measures. It does not understand. This is not a limitation that better sensors will resolve. Wearables are designed to capture biology continuously and passively, and they do so with increasing fidelity~\cite{piwek2016, liang2020}. But the psychological states that shape how we sleep, how much we move, and how well our bodies recover from stress are not directly encoded in heart rate or accelerometry. The past decade has seen substantial investment in longitudinal wearable studies of university students, a population at disproportionate risk for sleep disruption, physical inactivity, and mental health difficulties across the transition to and through higher education~\cite{lund2010, copeland2021, lipson2019}. These studies have revealed that sleep quality and physical activity deteriorate predictably across the academic year, that examination periods amplify this deterioration, and that individual trajectories vary considerably~\cite{wang2014, lund2010, galambos2009, nepal2024}. What wearable data cannot establish is what is driving those trajectories for a given student in a given week, and therefore what might usefully be done about it.

This gap has practical consequences for researchers and practitioners building passive sensing systems for student health. Without knowing whether a student's declining sleep efficiency reflects academic overload, relational distress, or financial worry, a wearable-derived alert is difficult to interpret and harder still to act on. Researchers have turned to complementary data streams like ecological momentary assessment, passive phone sensing, and social media language to capture the psychological context that wearable sensors cannot recover, each trading off participant burden, ecological validity, and interpretability differently~\cite{shiffman2008, harari2017, guntuku2017}. Yet brief naturalistic text collected as part of the study protocol itself has received comparatively little systematic attention as a complement to passive sensing, despite its potential to capture psychological context at 
minimal burden. 

Open-ended concern prompts can generate responses in the participant's own words at minimal burden, without the constraints of fixed-response scales, the noise of unsolicited social media data, or the effort of diary entries~\cite{zatzick2001}. Language is not merely a report of psychological state; it is an expression of it. The words a person reaches for when asked what concerns them, the emotional register they adopt, and the degree to which they elaborate or compress their response all reflect their psychological state at the moment of writing~\cite{pennebaker2003, tausczik2010}. Whether that signal lies in the topic of concern or in the affective register of its expression has not been directly tested. This distinction matters practically: a system that classifies concern topics will produce fundamentally different outputs, and require different infrastructure, than one that extracts emotional tone and valence from the same text.

We present a formative longitudinal study of $N=458$ university students tracked with Oura ring wearables across a full academic year (33 weeks, 3,610 person-waves, where each person-wave represents one participant's data from a single bimonthly survey cycle)\cite{price2023lemurs, bloomfield2024predictors, bloomfield2024predicting, fudolig2025collective}. Each bimonthly wave, participants responded to a brief open-ended concern prompt (``Of all the things that have happened in the last 2 weeks, what concerns you the most?''). The student from our opening vignette, for instance, might respond with nothing more than ``upcoming test,'' a response typical of the corpus in both length and content (see Table~\ref{tab:examples} for further examples). This generated a corpus of 3,073 concern-present responses with a median length of three words. We apply three NLP approaches---SEANCE dictionary features, RoBERTa-base embeddings, and MentalRoBERTa domain-adapted embeddings---and examine their associations with nine wearable-derived sleep and physical activity outcomes using a mixed-effects framework that isolates week-to-week variation from stable between-person differences.

\rev{The related literature reveals three specific gaps that this study addresses. First, while micro-EMA and voice-based self-report systems have demonstrated that low-burden input is feasible across device form factors, none have examined whether the linguistic content of ultra-brief open-ended 
responses associates with concurrent physiological outcomes in a longitudinal within-person design. Second, while language-based digital phenotyping has established associations between text and health outcomes, this work has focused on long-form or passively collected text and on between-person comparisons, leaving open the question of what signal remains in the kinds of ultra-brief responses that emerge when participants are not prompted to elaborate. Third, it remains untested whether the physiologically relevant signal in such text lies in topical content, what students are worried about or in the affective register in which concern is expressed, a distinction with direct practical consequences for system design. The following research questions address each gap in turn:}

\begin{itemize}
    \item \textbf{RQ1.} Does the linguistic content of brief naturalistic concern text associate with within-person variation in wearable-derived sleep and physical activity outcomes across a full academic year?
    \item \textbf{RQ2.} How do dictionary-based (SEANCE), general pretrained (RoBERTa-base), and domain-adapted (MentalRoBERTa) NLP approaches compare in their ability to explain variance in wearable outcomes from short longitudinal concern text?
    \item \textbf{RQ3.} What is the nature of the physiologically relevant signal in short response texts? Does it lie in topical content (what students are worried about), affective expression (how they express concern), or specific linguistic dimensions recoverable only through distributional representations?
\end{itemize}

\rev{The concern prompt used in this study is negatively valenced by construction, selectively surfacing worry and distress; our findings and contributions below should be read within that scope, and we return to what a positively framed prompt might add in our Limitations.}

\noindent We make the following contributions:

\begin{itemize}
    \item We demonstrate that brief naturalistic concern text associates with within-person variation in wearable-derived sleep and physical activity outcomes across a full academic year, even at a median response length of three words.
    \item We systematically compare dictionary-based (SEANCE), general pretrained (RoBERTa-base), and domain-adapted (MentalRoBERTa) NLP approaches for extracting signal from short longitudinal text, finding that general pretrained embeddings outperform or match domain-adapted models across most outcomes, including autonomic measures such as RMSSD.
    \item We show that emotional tone, not concern topic, carries the physiologically relevant signal, and that a substantial proportion of that signal is recoverable only through distributional representations and not by dictionary-based approaches.
\end{itemize}

\rev{The analyses proceed in four steps: documenting semester-level wearable trajectories, testing language-outcome associations for each NLP method, comparing variance explained across methods, and interpreting dominant embedding dimensions.}
\section{Related Work}
\label{sec:related}

Wearable devices enable continuous, passive measurement of sleep, physical activity, and heart rate variability at scale~\cite{piwek2016, smuck2021}, but the physiological signals they produce are difficult to interpret in isolation. Similar patterns may reflect stress, environmental demands, or
deliberate behavior change, and these distinctions are not recoverable from sensor data alone~\cite{mohr2017, huckvale2019}. Efforts to add psychological context through self-report methods such as ecological momentary assessment (EMA) improve interpretability but introduce participant burden and limit long-term scalability~\cite{shiffman2008, heron2017}. Language-based approaches offer a complementary path, but existing work tends to rely on long-form or passively collected text that is rarely aligned with concurrent wearable data in a temporally controlled manner~\cite{pennebaker2003, eichstaedt2018, cornet2018, jacobson2020}.

We review \rev{five} bodies of work in turn: wearable sensing and digital phenotyping, \rev{low-burden user input as a complement to passive sensing}, multimodal integration of wearable and subjective data, language-based digital phenotyping\rev{, and sensor-language models}, with attention to the tradeoffs between contextual richness and user burden, before positioning our approach as addressing these limitations.

\subsection{Wearable Sensing and Digital Phenotyping}
\label{sec:related-sensing}

Wearable device adoption now exceeds 59\% among smartphone owners across diverse socioeconomic groups~\cite{shandhi2024assessment,
chandrasekaran2020patterns}, and these signals are considered important indicators of mental health and overall functioning~\cite{gomes2023}. Their ability to passively capture longitudinal data has driven increased use in clinical research and digital health. Within the ubiquitous computing community specifically, foundational work has established passive sensing as a viable approach to digital phenotyping of mental health~\cite{benzeev2015},
longitudinal studies of college students have demonstrated associations between passively sensed behavioral features and depression outcomes across the academic year~\cite{chikersal2021}, and more recent cross-site work has examined how these signals generalize across student populations from different countries~\cite{meegahapola2023}.

Despite these strengths, wearable-derived signals are difficult to interpret in isolation. Changes in sleep, activity, or heart rate variability may reflect stress, environmental demands, or deliberate behavior change, and these distinctions are not directly observable from sensor data alone~\cite{mohr2017, huckvale2019}, limiting the ability to characterize mental health states or tailor interventions to individual needs. \rev{This interpretability problem is compounded by heterogeneity in how physiological-signal models are built and evaluated: a large-scale benchmark across 19 emotion-recognition datasets found that model performance varies systematically with collection protocol, device type, and labeling strategy, underscoring that dataset design choices materially shape what these models learn~\cite{singh2026feel}.}

\subsection{\rev{Low-Burden User Input as a Complement to Passive Sensing}}
\label{sec:related-input}

To address the lack of psychological context in passive sensing, many studies have incorporated self-report methods such as EMA, which capture in-the-moment reports of mood, stress, and behavior in close temporal proximity to experience~\cite{moskowitz2006ecological, shiffman2008, stone1994}. When combined with wearable data, these methods provide a more comprehensive view of within-person dynamics over time~\cite{trull2013, trull2020}, and EMA-based monitoring has been shown to support patient engagement and clinical decision-making~\cite{krohn2022}. EMA requires repeated prompts over extended periods, however, introducing participant burden, reducing adherence, and limiting long-term scalability~\cite{heron2017, wen2017}. Structured self-report measures further constrain how individuals can describe their experiences, potentially missing the cognitive and affective texture of what they are actually going through.

\rev{A body of work in the ubiquitous computing community has examined how to reduce this burden while preserving data quality. Micro-EMA ($\mu$EMA)
approaches deliver single-question prompts at a glance on a smartwatch, achieving higher compliance than multi-item EMA over shorter
deployments~\cite{le2024collecting}. Le et al.~\cite{le2024collecting} extended $\mu$EMA by introducing audio-based input, a short beep cues participants to verbally report their physical activity and posture state, demonstrating that speech enables open-ended responses at wrist level without the constraints of a small screen. A subsequent study examined the feasibility and utility of multimodal $\mu$EMA combining touch and voice modalities, finding that modality preference varies by task and context~\cite{le2025feasibility}. Beyond the wrist, Lim et al.~\cite{lim2024exploring} designed a context-aware ESM system using a
multimodal smart speaker that detects user context transitions to schedule surveys and allows responses via voice or touch; a four-week field study ($N = 20$) showed that context-triggered delivery increases compliance relative to random prompting. Kim et al.~\cite{kim2022mymove} built MyMove, a speech-based smartwatch application enabling older adults to label their physical activities in situ through verbal reports, pairing these labels with concurrent sensor data to support personalized activity tracking. Together, these works establish that brief, low-burden input is feasible across device form factors, and that the design of the input modality shapes what information users can express. However, existing systems have focused on structured behavioral self-reports rather than open-ended affective expression, and none have examined whether the linguistic content of ultra-brief naturalistic responses carries physiologically relevant signal in a longitudinal within-person design.}

Open-ended or ultra-brief text inputs represent a lower-burden alternative: they allow individuals to express experiences in their own words without the repeated prompting burden of EMA, the predefined constraints of fixed-response scales, or the effort of diary entries. Prior work suggests that integrating such text-based features with passive sensing improves performance relative to sensor-only approaches, supporting the view that these modalities capture distinct but complementary aspects of user experience~\cite{liu2022}. Whether very short naturalistic text collected within a study protocol---responses of a few words, delivered bimonthly---carries reliable and informative signal in relation to concurrent biometric data remains underexplored.

\subsection{Multimodal Integration of Wearable and Subjective Data}
\label{sec:related-integration}

A growing body of research has examined multimodal approaches that combine wearable data with self-report or other subjective inputs, with the potential to improve prediction of health outcomes and provide a more nuanced understanding of behavior and experience~\cite{cornet2018, jacobson2020}. Prior work has combined passive sensing with subjectively reported life events to examine how short-term experiences such as discrimination reshape day-to-day behavioral patterns in college students~\cite{sefidgar2019}, illustrating the value of aligning objective sensor streams with subjective context at a fine temporal grain. Existing methods often rely on coarse or infrequent measures or higher-burden reporting strategies, however, limiting their ability to capture fine-grained, temporally aligned within-person dynamics.

\subsection{Language-Based Digital Phenotyping}
\label{sec:related-language}

Natural language has been established as a source of psychological signal across a range of settings. Linguistic features extracted from social media posts have been used to detect depression~\cite{dechoudhury2013, reece2017forecasting, reece2017instagram}, predict hospitalization~\cite{eichstaedt2018}, and track population-level mental health~\cite{guntuku2017}; clinical transcripts and diary entries have been linked to symptom trajectories~\cite{resnik2015,nobles2018}; and typed messages and app usage logs have been associated with self-reported stress and wellbeing~\cite{wang2014, harari2017, xu2019}. Dictionary-based tools such as LIWC~\cite{tausczik2010} and SEANCE~\cite{crossley2017} operationalize this approach across psycholinguistic dimensions including affect, cognition, and social orientation, while embedding-based models capture semantic and contextual patterns that fixed lexicons cannot represent~\cite{mikolov2013, devlin2019}. This body of work has focused primarily on long-form or passively collected text and on between-person differences, however, with little attention to whether very short naturalistic text carries physiologically relevant signal in a longitudinal within-person design, or to which NLP approach is most appropriate when responses are brief and sparse.

General-purpose pretrained models such as RoBERTa~\cite{liu2019roberta} have become standard for downstream text tasks without task-specific tuning. Domain adaptation through continued pretraining on mental health corpora has produced variants such as MentalRoBERTa, trained on Reddit posts from mental health subreddits~\cite{ji2022mentalbert, stupinski2022quantifying}. Empirical evidence for the advantage of domain adaptation is mixed, however, particularly when the target text differs in genre, register, or length from the pretraining corpus, as is the case here where responses are brief and writers have not explicitly identified themselves diagnostically.

\rev{\subsection{Sensor-Language Models and the Direction of Integration}}
\label{sec:related-sensorlm}

\rev{A parallel line of work has pursued the inverse integration direction:
generating natural language descriptions \emph{from} sensor data rather than extracting signal from user-generated text. SensorLM~\cite{zhang2026sensorlm}, a sensor-language foundation model trained on over 59.7 million hours of multimodal wearable data from more than 103{,}000 participants, maps raw physiological signals to natural language descriptions through a hierarchical captioning pipeline, enabling zero-shot and few-shot recognition of health states from sensor data alone. This class of models illustrates the growing ambition to make wearable data linguistically interpretable at scale.}

\rev{The present work occupies a complementary position. Rather than generating language from sensors, we ask whether brief user-generated language, responses of a median three words to an open-ended affective prompt, carries detectable signal about concurrent sensor-derived outcomes within persons over time. Whereas sensor-language models require massive pretraining datasets and significant computational infrastructure, the approach examined here requires only a single bimonthly text prompt integrated into an existing study protocol. These two directions reflect different points in a design space: infrastructure-intensive language generation from sensors at one end, and lightweight affective capture from users at the other. The present study establishes what the lower end of that design space can recover.}
\rev{Complementary work has explored pairing physiological signals with textual descriptions in controlled settings: EEVR~\cite{singh2024eevr} 
introduced a VR-based dataset of EDA and PPG signals paired with qualitative textual descriptions of emotional experience, demonstrating that text supervision improves emotion recognition performance relative to signal-only baselines. More recently, the Personal Health Insights Agent (PHIA)~\cite{merrill2026phia} deployed an LLM agent that interprets longitudinal wearable data through multi-step reasoning and code generation to deliver personalized health insights, achieving 84\% accuracy on objective health questions. Together with SensorLM, these works illustrate the growing momentum toward making wearable data linguistically interpretable through large-scale model infrastructure.} \rev{Related efforts have examined complementary directions: AnnoSense~\cite{singh2025annosense} proposed a framework for collecting physiological emotion data in everyday settings using wearables and smartphones, and Mind Mapper~\cite{danry2026mindmapper} used wearable AI and LLMs to model behavioral patterns from everyday 
conversations. While these works differ from ours in scale, modality, and method, they collectively reflect a growing interest in integrating linguistic and physiological streams to support health monitoring and self-understanding.}

\rev{Together, these five bodies of work motivate the present study. The micro-EMA and voice-based self-report literature has established that low-burden input is feasible and that modality shapes expressiveness, but has not examined whether the linguistic content of ultra-brief open-ended responses associates with concurrent physiological outcomes in a longitudinal within-person design. The sensor-language model literature has established that language and wearable data can be aligned at scale when language is generated from sensors, but has not addressed whether user-generated affective text provides complementary signal that sensors cannot recover. The language-based digital phenotyping literature has established that linguistic features associate with health outcomes, but has focused on long-form or passively collected text and between-person comparisons. The present study addresses the intersection of these gaps.}

\section{Methodology}
\label{sec:methodology}

\begin{figure*}[t]
  \centering
  \includegraphics[width=0.8\textwidth]{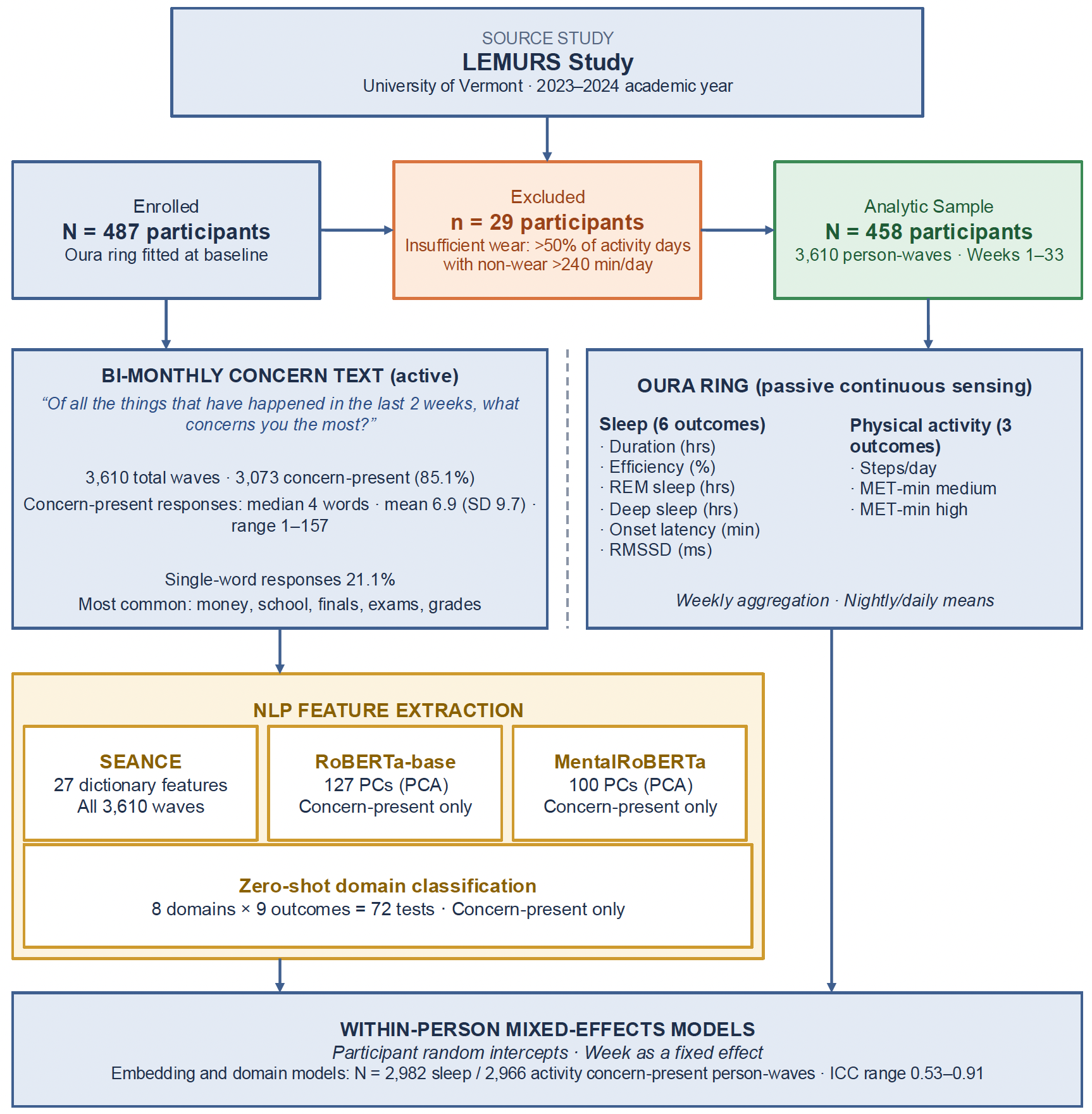}
  \caption{Study design and analytical pipeline. Participants from the LEMURS study (\cite{price2023lemurs};  NCT05841979) provided bi-monthly open-ended concern text and wore an Oura Ring continuously for 33 weeks. Concern text was processed through three NLP pipelines (SEANCE, RoBERTa-base, MentalRoBERTa) and zero-shot domain classification; Oura Ring data yielded nine wearable outcomes (six for sleep, three for physical activity). All pipelines fed within-person mixed-effects models with random intercepts per participant.}
  \Description{Diagram showing the study design, data collection, and analysis pipeline.}
  \label{fig:study-design}
\end{figure*}

\subsection{Study Design and Participants}
\label{sec:study-design}

This study is a secondary analysis of longitudinal data collected as part of Lived Experiences Measured Using Rings Study (LEMURS) \citep{price2023lemurs}, a Phase II parallel-group randomized controlled trial examining the efficacy of three wellbeing interventions for first-year university students. The parent study enrolled participants at \rev{the University of Vermont} during the 2023--2024 academic year and was approved by the University of Vermont Institutional Review Board (ClinicalTrials.gov: \rev{NCT05841979}). All participants provided written informed consent prior to enrollment. Participants were randomly assigned to one of four conditions: group-based therapy, a physical activity program, nature experiences, or a weekly assessment control condition. All participants wore an Oura ring continuously across the academic year and completed bimonthly self-report surveys via the study mobile application. \rev{The bimonthly interval was set by the parent study protocol to balance longitudinal coverage against participant burden across a 33-week academic year; the present analyses treat this interval as a design constraint and examine what signal remains detectable at this conservative sampling rate.} \rev{A shorter interval, such as weekly, would likely capture more transient affective fluctuations at the cost of greater participant burden and attrition risk over a full academic year, while a longer interval would further limit sensitivity to within-person change; the bimonthly interval represents a middle point in this tradeoff rather than an optimum established here.} \rev{The bimonthly concern prompt was designed to elicit a retrospective summary of psychological experience across the preceding two-week window rather than a momentary state, consistent with the post-traumatic concerns methodology from which it was adapted~\cite{zatzick2001}.} This retrospective framing is intentionally aligned with the bimonthly aggregation window applied to wearable outcomes, which average daily physiological measurements across the same two-week period. The design therefore does not assume that a single momentary report captures hourly physiological fluctuations; rather, both the linguistic predictor and the wearable outcomes reflect the same two-week window of experience. The present analyses treat all four arms as a single longitudinal observational cohort.

Of the 534 participants matched to wearable data, 487 met the minimum three-wave threshold for inclusion in the analytic sample; of these, 29 were excluded due to insufficient Oura ring wear time during activity monitoring (more than 50\% of activity days with non-wear exceeding 240 minutes per day). Sleep data for these 29 participants did not differ from the retained sample (mean efficiency 87.1\%, mean RMSSD 69.8 ms), but exclusion was applied consistently across all analyses. The analytic sample therefore comprised $N = 458$ participants who contributed valid wearable data across a minimum of three survey waves, yielding 3,610 person-waves. Data collection spanned the full 2023--2024 academic year (Weeks 1--33, Fall semester, winter break, and Spring semester). Participants contributed a mean of 7.9 waves (median $= 8$, range $= 3$--$11$). The sample was predominantly female (64.5\%), predominantly White (86.9\%), and had a mean age of 18.6 years ($SD = 0.6$). Full demographic characteristics are reported in Appendix~\ref{app:demographics}.

\subsection{Measures}
\label{sec:measures}

\subsubsection{Concern Text (Predictor)}
\label{sec:concern-text}

Each bimonthly wave, participants responded to a single open-ended question in an ecological momentary assessment delivered via the study mobile application: \textit{Of all the things that have happened in the last 2 weeks, what concerns you the most?} Responses were submitted as free-text entries. Out of the 3,610 person-waves, 3,073 (85.1\%) contained a concern text response; the remaining 537 (14.9\%) were concern-absent, comprising either blank entries (\rev{$n = 92$}) or responses expressing no current concern (e.g., ``Nothing,'' ``N/a,'' ``Not much''; \rev{$n = 445$}), and were coded accordingly (\texttt{concern\_present = 0}).

Responses were minimally preprocessed: leading and trailing whitespace was removed and blank or punctuation-only entries were excluded. \rev{Additionally, 47 spelling corrections were applied to the raw text, affecting 53 responses (e.g., ``sxhool'' corrected to ``school,'' ``strees'' to ``stress''); no further filtering was applied.} Responses expressing no concern (e.g., ``nothing,'' ``nothin,'' ``not much :)'') were coded as concern-absent (\texttt{concern\_present = 0}), consistent with the breakdown above. \rev{A small number of semantically equivalent responses (e.g., ``Nothing really,'' ``Nothin''; $n = 13$) did not match the coding list exactly and were retained as concern-present; we note this as a minor coding inconsistency rather than a deliberate distinction.} \rev{The median response length across all 3,610 person-waves, the sample used for SEANCE feature extraction, was 3 words (mean $= 6.0$, $SD = 9.2$, range $= 1$--$157$). Restricted to the 3,073 concern-present responses specifically, the median response length was 4 words (mean $= 6.9$, $SD = 9.73$).} Table~\ref{tab:examples} illustrates the breadth of topics and range of response lengths observed in the corpus. \rev{Single-word responses accounted for 32.0\% of all 3,610 waves and 21.1\% of the 3,073 concern-present responses specifically; responses of three words or fewer comprised 54.9\% of all waves and 47.1\% of concern-present responses,} reflecting the low-burden passive format of the prompt. Response brevity was not experimentally constrained: participants were free to write as much or as little as they chose. The median length of three words reflects naturalistic expression under an open-ended prompt rather than an imposed limit, making it a property of the corpus to be characterized rather than a design decision to be justified.

\begin{table}[ht]
\centering
\caption{Example Concern Text Responses Illustrating Breadth of Topics and Range of Response Lengths}
\label{tab:examples}
\begin{tabular}{lll}
\hline
Response & Words & Domain \\
\hline
\textit{Nothing} & 1 & No concern \\
\textit{Money} & 1 & Finances \\
\textit{Exams} & 1 & Academic \\
\textit{Situationship ended} & 2 & Relationships \\
\textit{My injury} & 2 & Physical health \\
\textit{Midterms coming up} & 3 & Academic \\
\textit{My grandfather is sick} & 4 & Physical health \\
\textit{Finding housing for next year} & 4 & Housing \\
\textit{My grades this semester} & 4 & Academic \\
\textit{Drama with friends and falling behind on work} & 8 & Relationships \\
\textit{My anxiety and depression are starting to kick in again} & 9 & Mental health \\
\textit{Boyfriend broke up with me; school is overwhelming} & 9 & Relationships \\
\textit{Feeling unprepared for finals; work piling up and feeling overwhelmed} & 11 & Academic \\
\textit{Struggling with the adjustment to college-level classes and exams} & 12 & Academic \\
\hline
\end{tabular}
\begin{tablenotes}
\small
\item \textit{Note.} Responses shown verbatim except the final two rows, which are paraphrased to protect participant privacy. Domain labels reflect zero-shot classification results. Response length ranges from 1 to 157 words in the full corpus (median = 3).
\end{tablenotes}
\end{table}

\subsubsection{Wearable Outcomes}
\label{sec:wearable-outcomes}

Heart rate variability (HRV) was operationalized as the root mean square of successive differences (RMSSD), a standard time-domain measure of cardiac autonomic function sensitive to psychological stress~\citep{thayer2012}. Nine wearable outcomes were examined: six sleep outcomes and three physical activity outcomes, as detailed in Table~\ref{tab:outcomes}.

\begin{table}[ht]
\centering
\caption{Wearable Outcomes Derived from Oura Ring Data}
\label{tab:outcomes}
\begin{tabular}{lll}
\hline
Outcome & Unit & Description \\
\hline
\multicolumn{3}{l}{\textit{Sleep (6 outcomes)}} \\
\quad Total duration & hours & Total time asleep per night \\
\quad Efficiency & \% & Proportion of time in bed spent asleep \\
\quad REM duration & hours & Time spent in REM sleep \\
\quad Deep duration & hours & Time spent in deep (slow-wave) sleep \\
\quad HRV & RMSSD (ms) & Cardiac autonomic function during sleep \\
\quad Onset latency & minutes & Time taken to fall asleep \\
\multicolumn{3}{l}{\textit{Physical activity (3 outcomes)}} \\
\quad Step count & steps/day & Mean daily step count \\
\quad MET-min medium & MET-min/day & Moderate-intensity activity volume \\
\quad MET-min high & MET-min/day & High-intensity activity volume \\
\hline
\end{tabular}
\begin{tablenotes}
\small
\item \textit{Note.} HRV = Heart Rate Variability. RMSSD = root mean square of successive differences. MET = metabolic equivalent of task. Onset latency is reported in minutes given its typical scale of 5--30 minutes; all other duration outcomes are in hours. Oura ring validity for sleep metrics has been established against polysomnography~\citep{dezambotti2019, altini2021} and for steps and energy expenditure against laboratory and free-living criterion measures~\citep{kristiansson2023}.
\end{tablenotes}
\end{table}

\subsection{NLP Feature Extraction}
\label{sec:nlp}

Three approaches to extracting linguistic features from concern text were compared, representing a progression from interpretable dictionary-based methods to general-purpose and domain-adapted pretrained language model embeddings. \rev{This progression was chosen to directly test whether the additional representational capacity of embedding-based models over named dictionary 
features produces meaningful gains in extracting physiologically relevant signal from brief text, and whether domain adaptation to mental health content provides further advantage over general pretraining.} All three approaches were applied to the 3,073 concern-present person-waves. To ensure an apples-to-apples comparison across methods, variance decomposition and dominance analyses were restricted to concern-present waves for all three approaches, with \texttt{concern\_present} omitted as a covariate since it is constant in this subset. SEANCE features were additionally available for the full 3,610 person-waves; the full-sample SEANCE models (with \texttt{concern\_present} as a covariate) are reported separately for completeness.

\subsubsection{SEANCE (Dictionary-Based Features)}
\label{sec:seance}

The Sentiment Analysis and Cognition Engine \citep[SEANCE;][]{crossley2017} was used to extract named linguistic features from each concern text response. SEANCE applies a suite of validated psycholinguistic dictionaries to compute feature scores reflecting emotional, cognitive, and social dimensions of text, including affect, sentiment, and value orientations~\citep{mohammad2013, bradley1999, susanto2022, hu2004, stone1966}. All features were normalized by word count to account for response length variation. SEANCE produced approximately 80 named feature dimensions per response, each corresponding to a labeled psychological or linguistic construct (e.g., \texttt{Sadness\_EmoLex}, \texttt{Academ\_GI}, \texttt{Tension/Stress\_GALC}, \texttt{Valence}, \texttt{Dominance}).

\subsubsection{RoBERTa-Base (General Pretrained Embeddings)}
\label{sec:roberta}

Sentence-level embeddings were extracted using RoBERTa-base \citep{liu2019roberta}, an optimized BERT pretraining approach trained on general English text corpora comprising BookCorpus, English Wikipedia, CC-News, OpenWebText, and Stories \citep{trinh2018}. Each concern text response was tokenized using the model's native byte-pair encoding tokenizer with a maximum sequence length of 64 tokens; responses exceeding this threshold were truncated. Token-level hidden states from the final layer were mean-pooled, weighted by the attention mask, to produce a single 768-dimensional sentence embedding per response. To reduce dimensionality prior to mixed-effects modeling, principal components analysis (PCA) was applied to the standardized embedding matrix across all concern-present responses. Components were retained until cumulative explained variance reached 80\%, yielding 127 principal components.

\subsubsection{MentalRoBERTa (Domain-Adapted Embeddings)}
\label{sec:mentalroberta}

A second set of embeddings was extracted using MentalRoBERTa \citep{ji2022mentalbert}, a variant of RoBERTa-base further pretrained on approximately 13 million mental health-related Reddit posts, included to test whether domain-specific pretraining improves extraction of psychologically relevant signal from brief concern text. The same extraction and dimensionality reduction procedure was applied as for RoBERTa-base, yielding 100 principal components explaining 80\% of embedding variance.

\subsubsection{Zero-Shot Domain Classification}
\label{sec:zeroshot}

Zero-shot domain classification was included to address RQ3---whether topical content carries physiologically relevant signal independently of affective features---and is therefore conceptually distinct from the three NLP methods compared in the variance decomposition. It is reported separately rather than included in the variance decomposition. The topical domain of each concern text response was characterized using zero-shot natural language inference \citep{yin2019} via the \texttt{facebook/bart-large-mnli} model. Each response was classified against nine candidate domain labels derived from frequency analysis of the full concern text corpus (see Table~\ref{tab:examples}): academic workload and exams, money and finances, physical health and illness, mental health and anxiety, relationships and social life, housing and living situation, future plans and career, general stress, and no concern. The model returned a full probability distribution over all nine domains for each response. These probability scores were entered as continuous predictors in within-person mixed-effects models of the same form as Equation~\ref{eq:model}, with $\text{Week}_z$ and $\text{nwords}_z$ as covariates, enabling direct assessment of whether topical domain explains variance in wearable outcomes above and beyond semester timing.

\subsection{Statistical Analysis}
\label{sec:stats}

\subsubsection{Analysis Plan}
\label{sec:analysis-plan}

\rev{The statistical analyses reported in this section address the three research questions introduced in Section~\ref{sec:intro}. Consistent with the formative and exploratory goals of this study (Section~\ref{sec:mixedeffects}), these analyses characterize the direction and magnitude of associations rather than test confirmatory, pre-registered predictions. Table~\ref{tab:analysis-plan} maps each research question to the analysis that addresses it. Prior to these analyses, the semester decline models (Section~\ref{sec:semester-decline}) establish whether wearable outcomes shift systematically across the academic year, independent of language content; this descriptive step motivates the inclusion of \texttt{Week\_z} as a covariate in all subsequent models and is not itself a test of RQ1--RQ3.}

\begin{table}[h]
\small
\begin{tabular}{p{0.06\linewidth} p{0.44\linewidth} p{0.40\linewidth}}
\toprule
\textbf{RQ} & \textbf{What the analysis targets} & \textbf{Primary analysis} \\
\midrule
RQ1 & Within-person language fluctuations vs.\ concurrent wearable outcomes & Within-person mixed-effects models (\S\ref{sec:mixedeffects}), applied separately to each NLP feature set and each of the nine wearable outcomes \\
RQ2 & Variance explained by each NLP method above semester timing & Variance decomposition (\S\ref{sec:variance-decomp}), comparing dictionary-based, general pretrained, and domain-adapted approaches on identical data \\
RQ3 & Whether signal lies in topical content or affective register & Zero-shot domain classification vs.\ affect features; dominance analysis (\S\ref{sec:dominance}); embedding interpretation (\S\ref{sec:embedding-interp}) \\
\bottomrule
\end{tabular}
\smallskip\noindent\small\textit{Note.} RQ = research question.
\caption{Analysis plan mapping research questions to their primary analysis.}
\label{tab:analysis-plan}
\end{table}

\subsubsection{Within-Person Mixed-Effects Models}
\label{sec:mixedeffects}

All primary analyses were conducted using within-person linear mixed-effects models with a random intercept for participant (\texttt{record\_id}), fitted using the \texttt{MixedLM} implementation in Python \texttt{statsmodels}. This specification partitions outcome variance into stable between-person differences, absorbed by the random intercept, and dynamic within-person fluctuations, which are the target of inference. Associations between bimonthly linguistic features and wearable outcomes therefore reflect wave-to-wave covariation within individuals rather than between-person correlations, a critical distinction given the high between-person stability of the outcomes (ICC range: 0.53--0.91).

For each NLP method and outcome combination, the fitted model took the form:

\begin{equation}
Y_{ij} = \beta_0 + \beta_1 \cdot \text{Language}_{ij} + \beta_2 \cdot \text{Week}_{z,ij} + \beta_3 \cdot \text{nwords}_{z,ij} + u_i + e_{ij}
\label{eq:model}
\end{equation}

\noindent where $Y_{ij}$ is the wearable outcome for participant $i$ at wave $j$; $\text{Language}_{ij}$ represents the set of NLP features for that method (SEANCE features, PCA components, or domain probability scores); $\text{Week}_{z,ij}$ is the $z$-scored academic week (indexing time); $\text{nwords}_{z,ij}$ is the $z$-scored word count of the concern text response; $u_i$ is the random intercept for participant $i$; and $e_{ij}$ is the residual error. SEANCE models were run on all 3,610 person-waves with \texttt{concern\_present} included as a binary covariate; embedding models were restricted to the 3,073 concern-present waves, with \texttt{concern\_present} constant and therefore omitted. Embedding and domain-classification models entered all components (or all eight domain probabilities) jointly, with semester week and a participant random intercept but without a word-count covariate; adding word count left results essentially unchanged.

\rev{Both \texttt{Week\_z} and \texttt{nwords\_z} were standardized globally across all participants and waves (i.e., using the grand mean and standard deviation of each variable computed over the full analytic sample), rather than within-person. This ensures that the fixed-effect coefficient for \texttt{Week\_z} reflects the within-person association between semester progression and outcomes across the shared academic calendar, and that \texttt{nwords\_z} adjusts for response length on a common scale. The linear mixed-effects specification assumes a linear relationship between linguistic features and wearable outcomes. This assumption is consistent with the formative and exploratory goals of this study, where the primary aim is to characterize the direction and magnitude of associations rather than to optimize predictive accuracy \cite{bolger2013intensive}. Non-linear relationships between affective language and physiological outcomes are plausible; restricted cubic spline analyses of emotional exhaustion, for instance, have demonstrated significant threshold and inflection effects on other health and performance outcomes rather than a simple linear dose-response \rev{\citep{zhu2026nonlinear}}, and whether comparable threshold effects hold for sleep and activity outcomes specifically remains an important direction for future work.}

\rev{Formally, for a variable $x$ with observations across all participants and waves, the standardized value is computed as:\begin{equation}
z_i = \frac{x_i - \bar{x}}{s_x}
\end{equation}
where $\bar{x}$ is the grand mean and $s_x$ is the standard deviation 
computed over the full analytic sample. The fitted model specification 
corresponds to:
\begin{equation*}
\texttt{outcome} \sim \texttt{feature\_z} + \texttt{concern\_present} + 
\texttt{nwords\_z} + \texttt{Week\_z} + (1 \mid \texttt{record\_id})
\end{equation*}
where \texttt{feature\_z} represents the standardized linguistic feature 
for a given NLP method and \texttt{(1 \textbar{} record\_id)} specifies a random intercept per participant.}

\rev{Models included a random intercept per participant but not random slopes. This choice reflects two considerations. First, the formative nature of this study prioritizes characterizing whether associations exist across the sample rather than modeling individual variability in those associations. Second, 
fitting random slopes for each of the 243 linguistic feature--outcome combinations tested would introduce substantial computational burden and convergence risk given the sparse within-person data structure (median 8 waves per participant). Future work with longer or more frequent sampling designs 
would be better positioned to estimate random slopes and examine how language--physiology associations vary across individuals.}

To manage Type I error across 243 tests (27 SEANCE features $\times$ 9 wearable outcomes), primary SEANCE features (\texttt{Negative\_EmoLex}, \texttt{Sadness\_EmoLex}, \texttt{Fear\_EmoLex}, \texttt{Valence}, \texttt{Arousal}, \texttt{Dominance}) were tested with Bonferroni correction (adjusted $\alpha = .0009$); exploratory features were tested with FDR correction (Benjamini-Hochberg). No associations survived either correction. Given the formative nature of this study, we report all uncorrected associations ($p < .05$) as hypothesis-generating findings, consistent with the goal of characterizing the signal landscape rather than making confirmatory claims. Readers should interpret these associations accordingly.

\subsubsection{Semester Decline Models}
\label{sec:semester-decline}

Prior to the primary NLP analyses, the longitudinal trajectory of each wearable outcome across the academic year was characterized using separate within-person mixed-effects models \rev{with standardized academic week ($\text{Week}_z$), \texttt{concern\_present}, and $\text{nwords}_z$ as fixed effects} and a random intercept for participant. Model-predicted values at Week 1 and Week 33 were extracted to characterize the direction and magnitude of semester-level change for each outcome\rev{, holding \texttt{concern\_present} and $\text{nwords}_z$ at their mean values}.

\subsubsection{Variance Decomposition}
\label{sec:variance-decomp}

To quantify the incremental contribution of linguistic features to outcome variance and enable direct comparison across the three NLP methods, a sequential block $R^2$ decomposition was conducted for each method-outcome combination. All block models for all three methods were restricted to the 3,073 concern-present waves, with \texttt{concern\_present} omitted from all formulas as it is constant in this subset. Four nested models were estimated per outcome: a null model with random intercept only, a covariates model adding $z$-scored word count, a semester model adding $z$-scored academic week, and a language model adding the full NLP feature set for that method. The incremental marginal $R^2$ at each block ($\Delta R^2$) was computed as the difference in marginal $R^2$ between successive models, where marginal $R^2$ reflects the proportion of total outcome variance attributable to fixed effects \citep{nakagawa2013}. This block structure allows direct quantification and comparison of the variance explained by linguistic features above and beyond semester timing across all three methods on identical data.

\subsubsection{Dominance Analysis}
\label{sec:dominance}

Dominance analysis \citep{budescu1993} was used to rank the relative contribution of individual linguistic features to explained outcome variance within each method. For each outcome, the average incremental $R^2$ contribution across all possible subsets of remaining predictors was computed, and predictors were ranked by this average contribution. For SEANCE, dominance analysis was applied to the full set of features retaining significant associations with each outcome. For embedding models, dominance analysis was restricted to the significant PCA modes for each outcome, capped at the ten components with the largest absolute regression coefficients to maintain computational tractability.

\subsubsection{Embedding Interpretation}
\label{sec:embedding-interp}

To support interpretable comparison between embedding-derived principal components and named SEANCE constructs, a three-part interpretation procedure was applied to each PC that was a significant predictor of at least one wearable outcome. Pearson correlations were computed between each significant PC and all SEANCE feature dimensions across the 3,073 concern-present responses; the strongest correlation was used to assign a named semantic label to the dimension. PCs were then correlated with zero-shot domain probability scores to identify topical alignment. Finally, the five concern text responses scoring highest and lowest on each PC were extracted as human-readable exemplars of what the dimension captures at its extremes. This triangulated procedure allowed embedding dimensions to be characterized in terms directly comparable to named SEANCE constructs, supporting qualitative cross-method comparison of what linguistic content drives associations with wearable outcomes.

\subsubsection{Software and Reproducibility}
\label{sec:software}

All analyses were conducted in Python 3.11 using the following libraries: the Hugging Face Transformers library for embedding extraction, \texttt{statsmodels} (\texttt{MixedLM}) for mixed-effects models, and \texttt{scikit-learn} and \texttt{scipy} for PCA and correlation analyses respectively. \rev{Analysis code is available at \url{https://github.com/data-tonye/brief-affective-text-wearables}. The repository contains code only; no participant data, derived outputs, or participant text are included. De-identified data sharing is subject to institutional review board restrictions; requests may be directed to the corresponding author.}
\section{Results}
\label{sec:results}

Results are organized around a single question: does the linguistic content of ultra-brief concern text associate with within-person variation in objectively measured sleep and physical activity outcomes, and if so, what kind of content carries the signal? We first characterize the corpus and document semester-level wearable trajectories, then report language-outcome associations for each NLP method, compare variance explained across methods, and interpret the dominant embedding dimensions.

\subsection{Concern Text Characteristics}
\label{sec:corpus}

Of the 3,610 person-waves in the analytic sample, 3,073 (85.1\%) contained a concern text response; the remaining 537 (14.9\%) were concern-absent. Among concern-present responses, word counts were highly right-skewed: the median response was 3 words (mean $= 6.0$, $SD = 9.2$, range $= 1$--$157$). Single-word responses accounted for 32.0\% of the corpus ($n = 1{,}154$) and responses of three words or fewer comprised 54.9\% ($n = 1{,}983$), reflecting the low-burden format of the prompt. Only 17.1\% of responses contained ten or more words ($n = 619$). The most common single-word responses were \textit{money} ($n = 111$), \textit{school} ($n = 80$), \textit{finals} ($n = 72$), \textit{exams} ($n = 65$), and \textit{grades} ($n = 48$), indicating that academic and financial concerns dominated the corpus.

SEANCE feature coverage was moderate across responses, consistent with the short response lengths. Descriptive statistics for all wearable outcomes are reported in Table~\ref{tab:descriptives}.

\subsection{Semester Decline in Wearable Outcomes}
\label{sec:semester}

Within-person mixed-effects models revealed significant semester-level decline across \rev{six of nine} wearable outcomes (Table~\ref{tab:semester}). RMSSD declined from a model-predicted 66.9 ms at Week 1 to 63.4 ms at Week 33 ($\beta = -1.03$, $SE = 0.17$, $p < .001$), representing a 5.2\% reduction in cardiac autonomic function across the academic year. Sleep efficiency declined from 88.1\% to 87.5\% ($\beta = -0.17$, $SE = 0.04$, $p < .001$). Sleep onset latency increased from 16.4 to 17.4 minutes ($\beta = +0.31$, $SE = 0.09$, $p < .001$), indicating students took progressively longer to fall asleep as the year advanced. Step count declined from a predicted 11,455 steps/day at Week 1 to 10,777 steps/day at Week 33 ($\beta = -200.1$, $SE = 36.4$, $p < .001$), and MET-min medium declined modestly ($\beta = -2.81$, $SE = 1.30$, $p = .031$). MET-min high also declined significantly ($\beta = -2.37$, $SE = 1.07$, $p = .026$).

\begin{table}[ht]
\centering
\caption{Descriptive Statistics for Wearable Outcomes Across All Person-Waves}
\label{tab:descriptives}
\begin{tabular}{lrrrrrrr}
\hline
Outcome & $N$ & Mean & $SD$ & Median & Min & Max \\
\hline
\multicolumn{7}{l}{\textit{Sleep outcomes}} \\
HRV/RMSSD (ms)            & 3,494 & 64.23     & 30.53    & 58.65     & 13.00    & 227.00    \\
Sleep efficiency (\%)     & 3,494 & 87.79     & 3.37     & 88.26     & 62.83    & 94.71     \\
Sleep duration (hrs)      & 3,494 & 7.33      & 0.68     & 7.35      & 2.70     & 9.92      \\
Deep sleep (hrs)          & 3,494 & 1.56      & 0.30     & 1.56      & 0.58     & 3.24      \\
REM sleep (hrs)           & 3,494 & 1.63      & 0.28     & 1.63      & 0.62     & 3.39      \\
Sleep onset latency (min) & 3,494 & 16.89     & 7.31     & 15.29     & 3.67     & 77.50     \\
\multicolumn{7}{l}{\textit{Physical activity outcomes}} \\
Steps/day                 & 3,475 & 11,069.51 & 3,529.05 & 10,618.79 & 2,347.50 & 28,950.71 \\
MET-min medium            & 3,475 & 233.69    & 108.90   & 217.33    & 9.75     & 887.71    \\
MET-min high              & 3,475 & 61.32     & 120.68   & 18.71     & 0.00     & 1,283.14  \\
\hline
\end{tabular}
\begin{tablenotes}
\small
\item \textit{Note.} All values are means of daily observations aggregated across a 14-day window per wave; decimal precision reflects averaging rather than instrument resolution. $N$ varies slightly by outcome due to missing wearable data. RMSSD = root mean square of successive differences. MET = metabolic equivalent of task.
\end{tablenotes}
\end{table}

\begin{table}[ht]
\centering
\caption{Within-Person Mixed-Effects Models: Semester Decline in Wearable Outcomes}
\label{tab:semester}
\begin{tabular}{lrrrrrl}
\hline
Outcome & Week 1 & Week 33 & $\beta$ & $SE$ & $p$ & Direction \\
\hline
\multicolumn{7}{l}{\textit{Sleep outcomes} ($N = 3{,}494$ from 455 participants)} \\
HRV/RMSSD (ms)            & 66.89     & 63.41     & $-1.026$ & 0.168 & $< .001$\textsuperscript{†} & Decline \\
Sleep efficiency (\%)     & 88.06     & 87.48     & $-0.172$ & 0.040 & $< .001$\textsuperscript{†} & Decline \\
Sleep duration (hrs)      & 7.32      & 7.32      & $-0.002$ & 0.008 & $= .798$                    & ---     \\
Deep sleep (hrs)          & 1.56      & 1.55      & $-0.003$ & 0.003 & $= .208$                    & ---     \\
REM sleep (hrs)           & 1.63      & 1.61      & $-0.006$ & 0.003 & $= .063$                    & ---     \\
Sleep onset latency (min) & 16.35     & 17.40     & $+0.310$ & 0.087 & $< .001$\textsuperscript{†} & Increase \\
\multicolumn{7}{l}{\textit{Physical activity outcomes} ($N = 3{,}475$ from 454 participants)} \\
Steps/day                 & 11,454.65 & 10,776.58 & $-200.093$ & 36.435 & $< .001$\textsuperscript{†} & Decline \\
MET-min medium            & 238.52    & 228.99    & $-2.811$ & 1.300 & $= .031$\textsuperscript{†} & Decline \\
MET-min high              & 64.65     & 56.62     & $-2.368$ & 1.065 & $= .026$\textsuperscript{†} & Decline \\
\hline
\end{tabular}
\begin{tablenotes}
\small
\item \textit{Note.} Models fitted on all available observations per outcome; per-outcome analytic $N$ ranges from 3,475 to 3,494 due to missing wearable data. $\beta$ = unstandardized coefficient for \texttt{Week\_z} (standardized week number across Weeks 1--33). $SE$ = standard error. Week 1 and Week 33 values are model-implied predicted means. \textsuperscript{†}Significant at $p < .05$. All models included a random intercept for participant.
\end{tablenotes}
\end{table}

\subsection{Within-Person Variance Structure}
\label{sec:icc}

ICCs confirmed substantial between-person stability across all outcomes (range: \rev{0.53--0.91}; Table~\ref{tab:variance}), indicating that \rev{53--91\%} of total outcome variance reflects stable individual differences. The remaining within-person variance, the target of all language-outcome analyses, ranged from \rev{9\% to 47\%}, confirming week-to-week fluctuation available for association testing. RMSSD's ICC of \rev{0.91} is a structural constraint worth noting: the majority of cardiac autonomic variation in this sample is stable rather than state-dependent, limiting the available signal for any weekly predictor to explain.

\subsection{SEANCE: Dictionary-Based Feature Associations}
\label{sec:seance-results}

Bivariate within-person mixed-effects models yielded 21 nominally significant associations ($p < .05$, uncorrected) out of 243 tests (8.6\%), spanning seven of nine outcomes; no associations survived Bonferroni or FDR correction. No significant associations were found for total sleep duration or REM sleep. Results are reported in Table~\ref{tab:seance}.

\paragraph{Physical activity.} Forceful language (\texttt{Strong\_GI}) showed the strongest and most consistent associations, with weeks of stronger, more assertive concern language associated with fewer steps/day and lower MET-min medium. Power-oriented language (\texttt{Powtot\_Lasswell}) showed similar negative associations with both activity outcomes, while work-oriented language (\texttt{Work\_GI}) was positively associated with steps/day, suggesting that task-focused rather than forceful concern framing was associated with higher activity.

\paragraph{Sleep onset latency.} Higher SenticNet sensitivity scores were associated with shorter onset latency, while anxiety and anger language were associated with longer onset latency. The negative association of academic language (\texttt{Academ\_GI}) with onset latency is counterintuitive and should be interpreted with caution given its proximity to the significance threshold.

\paragraph{Sleep efficiency.} Sensitivity, work-oriented language, and academic language were each positively associated with sleep efficiency, suggesting that task-focused concern framing was associated with modestly better sleep consolidation.

\paragraph{RMSSD} Four ANEW-derived features---dominance, valence, arousal, and academic language---were each positively associated with cardiac autonomic function, indicating that weeks with higher emotional valence and more academic concern framing were associated with better autonomic recovery.

\paragraph{Deep sleep.} Sadness language (\texttt{Sadness\_EmoLex}) was negatively associated with deep sleep duration, with higher sadness in concern text associated with less deep sleep that week.

\paragraph{Dominance analysis.} \rev{\texttt{Week\_z} was the leading predictor for five of six outcomes: RMSSD (55.6\%), sleep efficiency, sleep onset latency, and both physical activity outcomes. Among linguistic features, \texttt{Strong\_GI} was the dominant predictor for MET-min medium, though for steps/day it ranked second to \texttt{Week\_z}. For RMSSD, \texttt{Week\_z} led (55.6\%), followed by academic language (\texttt{Academ\_GI}; 14.6\%), valence (9.6\%), sadness (9.1\%), and fear (2.6\%). For sleep onset latency, academic language and semester timing were closely matched (38.8\% vs.\ 36.2\%, respectively), rather than one clearly surpassing the other.}

\begin{table}[ht]
\centering
\caption{Significant SEANCE Feature Associations with Wearable Outcomes ($p < .05$)}
\label{tab:seance}
\begin{tabular}{llrrr}
\hline
Feature & Construct & $\beta$ & $SE$ & $p$ \\
\hline
\multicolumn{5}{l}{\textit{Physical activity --- steps/day}} \\
\texttt{Strong\_GI}       & Forceful/assertive language & $-133.50$ & 38.71 & $< .001$ \\
\texttt{Afftot\_Lasswell} & Affective social language   & $-94.79$  & 38.32 & $= .013$ \\
\texttt{Powtot\_Lasswell} & Power-oriented language     & $-87.12$  & 36.83 & $= .018$ \\
\texttt{Work\_GI}         & Work/task-focused language  & $+83.66$  & 42.41 & $= .049$ \\
\multicolumn{5}{l}{\textit{Physical activity --- MET-min medium}} \\
\texttt{Strong\_GI}       & Forceful/assertive language & $-4.19$   & 1.38  & $= .002$ \\
\texttt{Powtot\_Lasswell} & Power-oriented language     & $-2.78$   & 1.31  & $= .034$ \\
\texttt{pleasantness}     & Pleasantness (SenticNet)    & $+2.70$   & 1.34  & $= .044$ \\
\texttt{polarity}         & Polarity (SenticNet)        & $+2.77$   & 1.40  & $= .049$ \\
\multicolumn{5}{l}{\textit{Physical activity --- MET-min high}} \\
\texttt{sensitivity}      & Sensitivity (SenticNet)     & $+2.31$   & 1.09  & $= .034$ \\
\multicolumn{5}{l}{\textit{Sleep efficiency}} \\
\texttt{sensitivity}      & Sensitivity (SenticNet)     & $+0.12$   & 0.04  & $= .004$ \\
\texttt{Work\_GI}         & Work/task-focused language  & $+0.10$   & 0.05  & $= .030$ \\
\texttt{Academ\_GI}       & Academic language           & $+0.09$   & 0.05  & $= .048$ \\
\multicolumn{5}{l}{\textit{Sleep onset latency}} \\
\texttt{sensitivity}      & Sensitivity (SenticNet)     & $-0.29$   & 0.09  & $= .001$ \\
\texttt{Academ\_GI}       & Academic language           & $-0.25$   & 0.10  & $= .016$ \\
\texttt{Anxiety\_GALC}    & Anxiety language (GALC)     & $+0.21$   & 0.09  & $= .026$ \\
\texttt{Anger\_EmoLex}    & Anger language (EmoLex)     & $+0.21$   & 0.10  & $= .039$ \\
\multicolumn{5}{l}{\textit{RMSSD (cardiac autonomic function)}} \\
\texttt{Dominance}        & Dominance (ANEW)            & $+0.44$   & 0.20  & $= .029$ \\
\texttt{Valence}          & Valence (ANEW)              & $+0.41$   & 0.20  & $= .040$ \\
\texttt{Academ\_GI}       & Academic language           & $+0.41$   & 0.20  & $= .041$ \\
\texttt{Arousal}          & Arousal (ANEW)              & $+0.41$   & 0.20  & $= .042$ \\
\multicolumn{5}{l}{\textit{Deep sleep}} \\
\texttt{Sadness\_EmoLex}  & Sadness language (EmoLex)   & $-0.008$  & 0.003 & $= .010$ \\
\hline
\end{tabular}
\begin{tablenotes}
\small
\item \textit{Note.} All models estimated via within-person mixed-effects regression: outcome $\sim$ \texttt{feature\_z} + \texttt{concern\_present} + \texttt{nwords\_z} + \texttt{Week\_z} + (1 $|$ \texttt{record\_id}). $N = 3{,}494$ person-waves from 455 participants (sleep outcomes) / $N = 3{,}475$ person-waves from 454 participants (activity outcomes). Features are standardized ($z$-scored). $\beta$ = unstandardized coefficient for the standardized linguistic feature. Primary features (\texttt{Negative\_EmoLex}, \texttt{Sadness\_EmoLex}, \texttt{Fear\_EmoLex}, \texttt{Valence}, \texttt{Arousal}, \texttt{Dominance}) tested with Bonferroni correction; exploratory features tested with FDR (Benjamini-Hochberg) correction. $p$-values shown are uncorrected; 21 of 243 tests were significant at $p < .05$. ANEW = Affective Norms for English Words. EmoLex = NRC Emotion Lexicon. GI = General Inquirer. GALC = Geneva Affective Label Coder. SenticNet = SenticNet 5.0.
\end{tablenotes}
\end{table}

\subsection{RoBERTa-Base Embedding Associations}
\label{sec:roberta-results}

Within-person mixed-effects models testing associations between RoBERTa-base PCA components and wearable outcomes yielded significant associations across all nine outcomes. Full results are reported in Table~\ref{tab:roberta-full}.

Several dimensions spanned multiple outcomes simultaneously. RoBERTa PC34 was negatively associated with sleep efficiency ($\beta = -0.056$, $p = .008$) and positively associated with onset latency ($\beta = +0.125$, $p = .007$), indicating a single linguistic dimension that captures worse sleep consolidation and longer time to fall asleep concurrently. PC39 implicated the same two outcomes, with signs reversed relative to PC34 (an artifact of arbitrary PC polarity, not opposing effects), further supporting the presence of a stable
emotional-linguistic dimension that covaries with disrupted sleep within students above and beyond semester timing. PC117 was associated with four outcomes simultaneously---less sleep duration and REM sleep, but more steps/day and MET-min medium---suggesting a dimension capturing high-engagement weeks that trade sleep quantity for physical activity. PC30 showed broad negative associations with sleep efficiency, steps/day, MET-min medium, and MET-min high simultaneously, suggesting a dimension capturing generalized low-energy concern framing that depresses both sleep quality and physical activity in the same week.

For physical activity, RoBERTa PC1---the dimension most strongly anchored by academic language (\texttt{Academ\_GI}, $r = +0.44$), with single-word academic responses at the positive pole and emotional or relational text at the negative pole---was negatively associated with steps/day ($\beta = -13.68$, $p = .036$). This replicates the SEANCE \texttt{Academ\_GI} finding: weeks dominated by academic concern framing were associated with lower physical activity within students, above and beyond semester timing.

\subsection{MentalRoBERTa Embedding Associations}
\label{sec:mentalroberta-results}

Within-person mixed-effects models testing associations between MentalRoBERTa PCA components and wearable outcomes yielded significant associations across all nine outcomes, with fewer unique significant PCs than RoBERTa-base (48 vs.\ 59) but stronger cross-outcome interpretability for mental health and emotional exhaustion dimensions. Full results are reported in Appendix Table~\ref{tab:mentalroberta-full}.

MentalRoBERTa PC1 replicated the dominant RoBERTa-base finding. Anchored by academic language (\texttt{Academ\_GI}, $r = +0.40$) with single-word academic responses at the positive pole and emotionally elaborated relational text at the negative pole, PC1 was negatively associated with steps/day and MET-min medium. The near-identical structure of RoBERTa PC1 and MentalRoBERTa PC1, both capturing the same academic-versus-emotional axis as the first principal component of their respective embedding spaces, confirms that academic concern framing is the dominant axis of variation in this corpus, independently recovered by two differently trained models.

\rev{One dimension highlighted a pattern unique to either model. PC7, anchored by negative emotion language (\texttt{Negative\_EmoLex}, $r = +0.36$), was negatively associated with both sleep efficiency ($\beta = -0.024$, $p = .049$) and RMSSD ($\beta = -0.137$, $p = .010$) simultaneously; RoBERTa PC7 shows the same dual association, so this pattern is not unique to the domain-adapted model.} PC31 was the most outcome-spanning dimension in either model, showing significant associations with sleep efficiency, REM sleep, RMSSD, and MET-min high simultaneously, with academic activity language at the positive pole and relational distress at the negative pole. This pattern was not recovered by RoBERTa.

RoBERTa's broader significant associations across activity outcomes suggest that general pretraining captures physical activity-relevant language more effectively than domain-adapted pretraining, likely because Reddit mental health discourse underrepresents the casual academic and physical activity language that characterizes brief concern text in a student population.

\subsection{Variance Decomposition: Three-Way Comparison}
\label{sec:variance}

To enable a direct apples-to-apples comparison across the three NLP methods, all variance decomposition models were restricted to the 3,073 concern-present waves, with \texttt{concern\_present} omitted as a constant covariate. Results are reported in Table~\ref{tab:variance}.

Across all outcomes, between-person stability accounted for the largest share of total outcome variance (ICC: 0.53--0.91), with semester timing (\texttt{Week\_z}) explaining a modest but consistent increment above covariates. \rev{Both embedding blocks showed a larger language-block $\Delta R^2$ than SEANCE across all nine outcomes. The embedding blocks contain substantially more predictors (100--127 principal components) than the SEANCE block (seven pre-selected features), and these $\Delta R^2$ values are estimated in-sample; taken together, this indicates that embedding blocks fit the data better under this specification, rather than establishing that embeddings recover more physiologically relevant signal than dictionary-based features.} \rev{RoBERTa-base showed a numerically larger $\Delta R^2$ than MentalRoBERTa in eight of nine outcomes, including RMSSD; MentalRoBERTa was marginally ahead only for MET-min high, by a difference of .0004. Bootstrap confidence intervals for the RoBERTa--MentalRoBERTa difference included zero for steps/day and MET-min high, so these two outcomes are better described as comparable between models than as a clear advantage for either.} \rev{Language block $\Delta R^2$ values ranged from .0005 (SEANCE, REM sleep) to .0370 (RoBERTa, MET-min medium) across all three methods, and from .0047 to .0370 across the two embedding models alone,} reflecting the combined effect of high between-person ICC, brief concern text, and the conservative within-person modeling approach. Within this constrained variance structure, even small language block increments represent notable proportions of the available within-person signal.

\begin{table}[ht]
\centering
\caption{Variance Decomposition: Language Block $\Delta R^2$ Across Three NLP Methods}
\label{tab:variance}
\rev{
\begin{tabular}{lcccccl}
\toprule
Outcome & ICC & $\Delta R^2$ Sem. & $\Delta R^2$ SEANCE & $\Delta R^2$ RoBERTa & \shortstack{$\Delta R^2$\\MentalRoBERTa} & Best \\
\midrule
\multicolumn{7}{l}{\textit{Sleep outcomes}} \\
Sleep duration (hrs)      & .534 & .0001 & .0010 & \textbf{.0250} & .0208 & RoBERTa \\
Sleep efficiency (\%)     & .578 & .0023 & .0019 & \textbf{.0281} & .0231 & RoBERTa \\
REM sleep (hrs)           & .648 & .0005 & .0005 & \textbf{.0195} & .0142 & RoBERTa \\
Deep sleep (hrs)          & .753 & .0001 & .0011 & \textbf{.0157} & .0114 & RoBERTa \\
HRV/RMSSD (ms)            & .913 & .0009 & .0007 & \textbf{.0055} & .0047 & RoBERTa \\
Sleep onset latency (min) & .573 & .0015 & .0024 & \textbf{.0239} & .0176 & RoBERTa \\
\multicolumn{7}{l}{\textit{Physical activity outcomes}} \\
Steps/day                 & .685 & .0033 & .0027 & .0324 & .0284 & Comparable \\
MET-min medium            & .568 & .0006 & .0032 & \textbf{.0370} & .0325 & RoBERTa \\
MET-min high               & .756 & .0010 & .0010 & .0144 & .0148 & Comparable \\
\bottomrule
\end{tabular}
}
\smallskip
\noindent\small\raggedright\textit{Note.} $\Delta R^2$ values computed on concern-present rows only ($N = 2{,}982$ sleep; $N = 2{,}966$ activity); ICC computed on all 3,610 person-waves with an outcome. $\Delta R^2$ Sem.\ = incremental marginal $R^2$ for \texttt{Week\_z} above covariates. $\Delta R^2$ SEANCE = incremental marginal $R^2$ for seven pre-selected SEANCE features above \texttt{Week\_z}; $\Delta R^2$ RoBERTa and MentalRoBERTa = incremental marginal $R^2$ for the full set of principal components (127 and 100, respectively) above \texttt{Week\_z}. Best = embedding model with the larger point estimate; outcomes marked ``Comparable'' had bootstrap confidence intervals for the RoBERTa--MentalRoBERTa difference that included zero. Marginal $R^2$ computed using the \citet{nakagawa2013} method.\normalsize
\end{table}

\subsection{Dominance Analysis}
\label{sec:dominance-results}

Dominance analysis ranked predictors by their average incremental $R^2$ contribution across all possible model subsets. Full embedding dominance results are reported in Appendix~\ref{app:embedding-dominance}. Among SEANCE features, \texttt{Strong\_GI} was the single most important predictor of MET-min medium across all predictors including semester timing, and \texttt{Academ\_GI} was the primary driver of sleep onset latency. \rev{For both embedding models, individual PCs accounted for 15--37\% of explained $R^2$ for their dominant outcomes.}

\subsection{Embedding Interpretation}
\label{sec:interpretation}

Applying the three-part interpretation procedure described in Section~\ref{sec:embedding-interp} yielded interpretable semantic labels for 68\% of significant RoBERTa PCs and 83\% of significant MentalRoBERTa PCs. The remainder showed weak correlations with all SEANCE features ($|r| < 0.10$) and are designated as weak/unclear signal dimensions. Key dimensions are described below.

\begin{figure*}[htbp]
  \centering
  \includegraphics[width=1.0\textwidth]{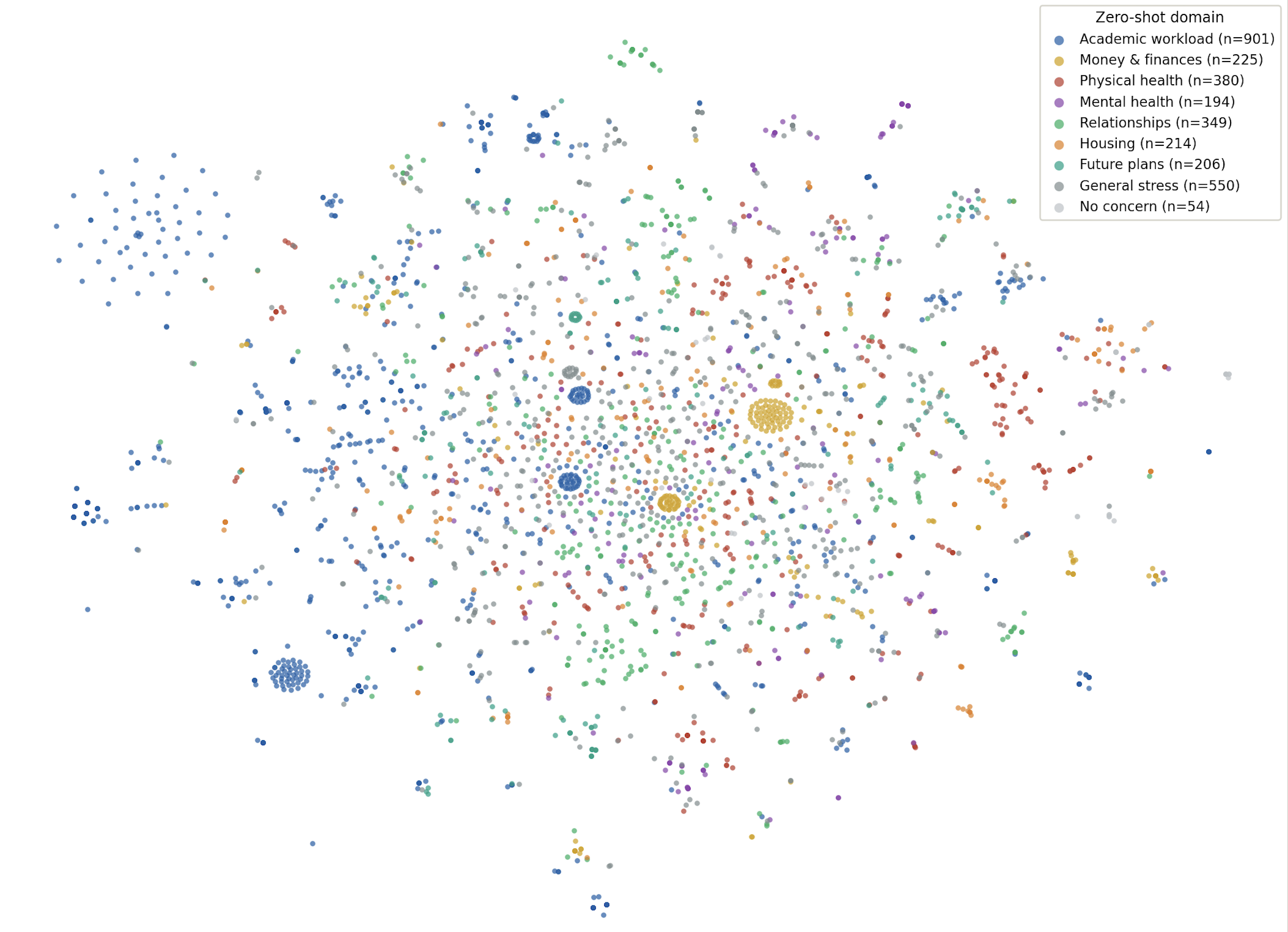}
  \Description{t-SNE projection of RoBERTa-base embeddings for all 3,073 concern-present student-weeks, colored by zero-shot domain classification. Academic workload responses form a spatially coherent cluster; single-word responses produce dense stacking. Domain labels reflect the highest-confidence zero-shot classification per response.}
  \caption{t-SNE projection of RoBERTa-base embeddings for all 3,073 concern-present student-weeks (perplexity = 40), colored by zero-shot domain classification. Academic workload responses ($n = 901$) form a spatially coherent cluster in the left region of the embedding space; single-word responses (e.g., \textit{money}, \textit{grades}) produce dense stacking visible in the center. Domain labels reflect the highest-confidence zero-shot classification per response.}
  \label{fig:tsne-global}
\end{figure*}

\paragraph{Academic pressure as the primary activity-suppressing dimension.} The first principal component of both embedding models was anchored by academic language (\texttt{Academ\_GI}: RoBERTa PC1 $r = +0.44$, MentalRoBERTa PC1 $r = +0.40$) and negatively aligned with general stress domain scores. Both PC1s showed near-identical loading structures---single-word academic responses (\textit{school}, \textit{finals}, \textit{chem}, \textit{hw}) at the positive pole and emotionally elaborated relational or distress language at the negative pole---and both were negatively associated with steps/day; only MentalRoBERTa PC1
reached significance for MET-min medium ($p = .045$), while RoBERTa PC1 did not ($p = .062$). The near-perfect replication of this dimension as the primary component across two independently trained embedding models confirms that academic concern framing is the dominant axis of variation in this corpus and is robustly associated with reduced physical activity within students above and beyond semester timing.

\paragraph{Emotional tone as the primary sleep-relevant dimension.} Across both models, the dimensions most consistently associated with sleep outcomes were characterized by emotional affect rather than topical content. RoBERTa PC34, showing significant associations with both sleep efficiency (negative) and sleep onset latency (positive), indicates a single linguistic dimension that simultaneously captures worse sleep consolidation and longer time to fall asleep, above and beyond semester timing. RoBERTa PC39 implicated the same two outcomes, with signs reversed relative to PC34 (an artifact of arbitrary PC polarity, not opposing effects), further supporting the presence of a stable emotional-linguistic dimension that covaries with disrupted sleep within students. MentalRoBERTa PC7, anchored by negative emotion language (\texttt{Negative\_EmoLex}: $r = +0.36$) and characterized by emotional exhaustion language at the positive pole (direct participant quotes: \textit{just feeling kinda down}, \textit{feeling tired all the time}, \textit{felt really down and sad a lot}, \textit{felt emotionally tired}), was negatively associated with both sleep efficiency and RMSSD simultaneously, suggesting that the specific register of emotional exhaustion language captured by the domain-adapted model carries signal for both sleep quality and cardiac autonomic recovery.

\begin{figure*}[htbp]
  \centering
  \includegraphics[width=\textwidth]{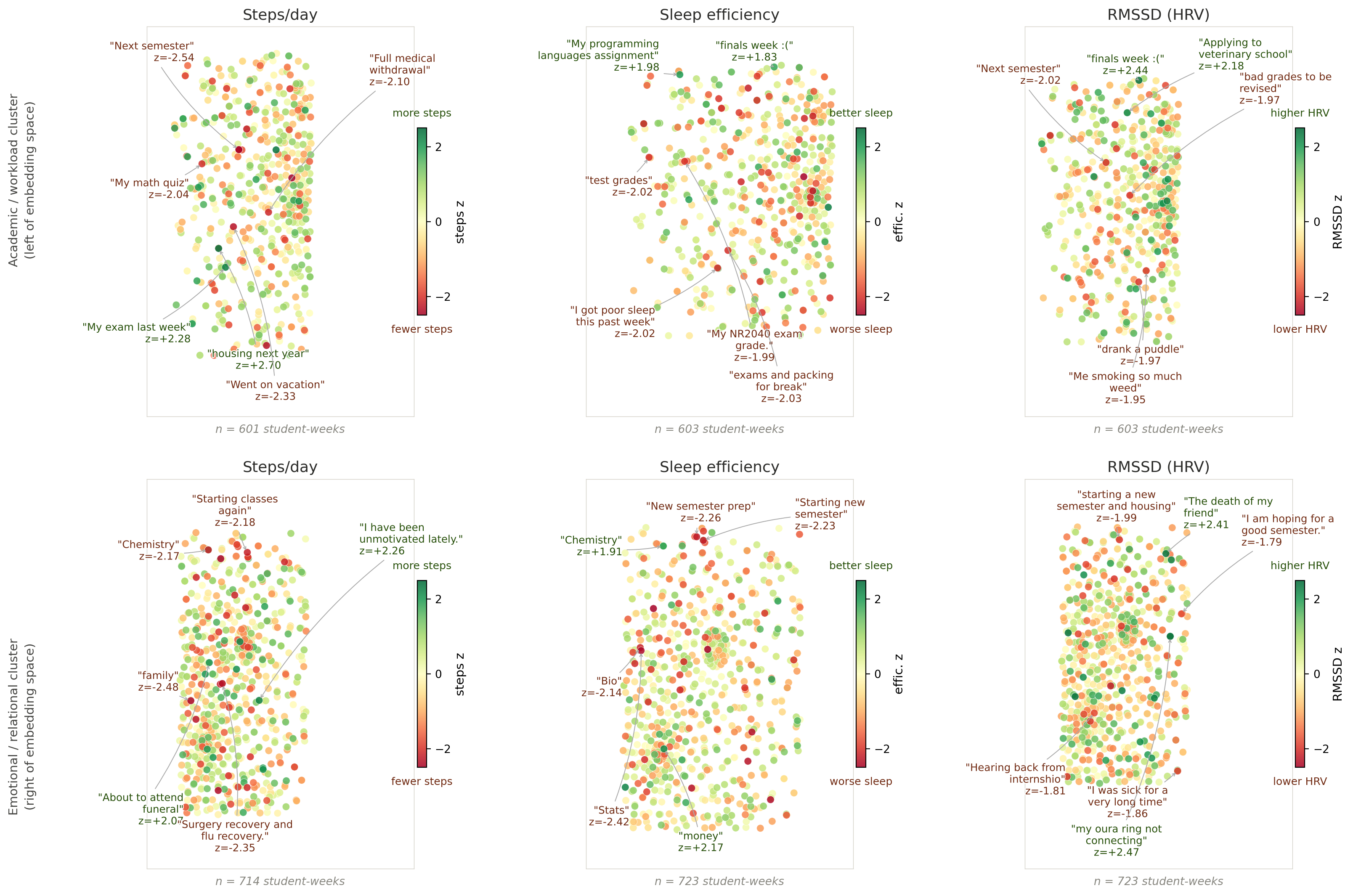}
  \caption{Within-person wearable outcomes across two zoomed semantic neighborhoods of the RoBERTa-base embedding space (axes are unlabeled t-SNE latent dimensions; see Figure~\ref{fig:tsne-global}). Each dot represents one student-week, colored by within-person $z$-scored outcome (green\,=\,above the student's own weekly average; red\,=\,below). Top row: academic/workload cluster. Bottom row: emotional/relational cluster, encompassing relationships, mental health, housing, and general stress domains. Annotated examples show concern texts at the extremes of the outcome distribution within each cluster (4 low-scoring\,:\,2 high-scoring texts per panel). All five outcomes are shown in Supplementary Figure~\ref{fig:supp-all-outcomes}.}
  \Description{Scatterplot of student-week data points in two clusters, colored by outcome, with annotated example texts.}
  \label{fig:zoomed-outcomes}
\end{figure*}

\paragraph{Topical content does not carry the signal.} Zero-shot domain classification produced largely null results across 72 tests. One association survived FDR correction: the physical health domain was negatively associated with sleep efficiency ($\beta = -1.79$, $SE = 0.46$, $p_{\text{fdr}} = .008$), consistent with students who are physically unwell sleeping less efficiently. \rev{This narrow exception aside, topical content did not independently covary with wearable outcomes within persons. Affective dimensions across all three NLP methods showed more associations with outcomes than topic did, though none of the SEANCE affect associations survived correction for multiple comparisons; this pattern is suggestive rather than confirmatory.} Full FDR-corrected results across all 72 domain; outcome combinations are reported in Appendix~\ref{app:domain}.

\paragraph{Dimensions with broad outcome coverage.} \rev{MentalRoBERTa PC31 was among the most outcome-spanning dimensions in either model, associating with sleep efficiency, REM sleep, RMSSD, and MET-min high simultaneously, though its dominance share for RMSSD is 18\%. RoBERTa PC30 and PC117 span a comparable number of outcomes (four each), so this breadth is not unique to the domain-adapted model. This comparison is limited to the top-10 significant PCs retained for dominance analysis in each model.} Weeks framed around academic activity were associated with better sleep quality and autonomic recovery; weeks framed around relational loss were associated with worse outcomes across multiple domains.

\begin{figure*}[htbp]
  \centering
  \includegraphics[width=\textwidth]{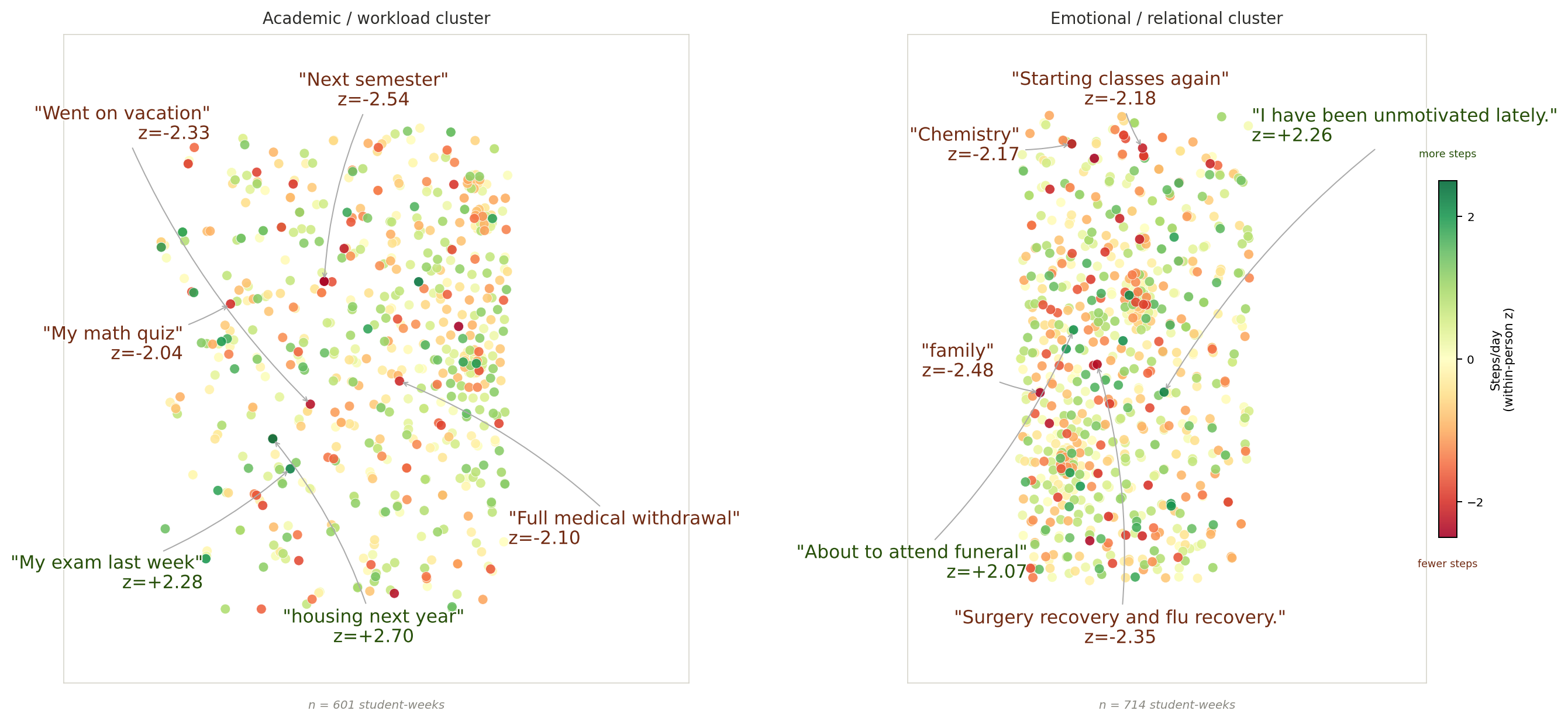}
   \caption{Daily step count (within-person $z$-score) across the academic/workload and emotional/relational semantic
  neighborhoods of the RoBERTa-base embedding space. Each dot represents one student-week; color encodes the within-person
  $z$-scored step count (green\,=\,above average;  red\,=\,below). Within each panel, the four lowest-activity and two highest-activity weeks are annotated (a 4:2 sampling of the extremes, not a summary of the panel's overall color balance, which is roughly even in both clusters: academic cluster, $n = 601$ student-weeks, 40\% green vs.\ 37\% red; emotional cluster, $n = 714$, 38\% green vs.\ 38\% red). The annotated low-activity weeks in the academic cluster reflect academic pressure, illness, and recovery, while the high-activity weeks reflect housing search and post-exam relief. This spatial contrast is consistent with RoBERTa's first principal component (Academ\_GI $r = +0.44$), which anchors single-word academic
  responses at one pole and emotionally elaborated relational text at the other and is negatively associated with steps/day ($\beta = -13.68$, $p = .036$; Section~\ref{sec:roberta-results}).}
  \Description{Scatter plot comparing two clusters of student-weeks positioned in RoBERTa embedding space, each dot colored by within-person step-count z-score, green for above average and red for below. The academic/workload cluster (n=601) and emotional/relational cluster (n=714) both show a roughly even mix of green and red dots, with no visible spatial separation by activity level. A small number of annotated points mark the four lowest-activity and two highest-activity weeks in each cluster, labeled with brief concern-text excerpts illustrating academic pressure, illness, and recovery (low-activity) versus housing search and post-exam relief (high-activity).}
  \label{fig:steps-panel}
\end{figure*}

\paragraph{Dimensions not recovered by named SEANCE features.} Thirty-two percent of significant RoBERTa PCs and 17\% of MentalRoBERTa PCs showed weak correlations with all SEANCE features ($|r| < 0.10$), designated as weak/unclear signal. These dimensions likely capture subtle pragmatic features---hedging, brevity as a stance, emotional minimization---not lexically indexed by dictionary approaches, pointing to a category of health-relevant linguistic signal recoverable only through distributional representations.
\section{Discussion}
\label{sec:discussion}

\subsection{Key Findings}
\label{sec:disc-findings}

Brief naturalistic concern text carries detectable linguistic signal about concurrent wearable-derived health outcomes above and beyond semester timing, even at a median response length of \rev{four words among concern-present responses}. Prior work incorporating text into health sensing has relied on longer or richer sources such as social media posts, clinical interviews, or multi-sentence EMA responses~\citep{dechoudhury2013, coppersmith2014, guntuku2017}. The present results suggest that a single ultra-brief bimonthly question yields linguistically informative responses, establishing a lower bound on what passive affective text sensing can achieve and positioning brief concern text as a practical complement to the psychological opacity of wearable sensing~\citep{mohr2017}.

\rev{The signal appears concentrated in emotional tone rather than topical content, though the evidence for this is suggestive rather than conclusive. Zero-shot domain classification produced one association surviving FDR correction (physical health $\rightarrow$ sleep efficiency, $p_{fdr} = .008$);
no other domain-outcome association survived correction (\S~\ref{sec:interpretation}, Table \ref{tab:domain-full}), and affective dimensions across all three NLP methods showed more associations with outcomes than topic did; however, none of the SEANCE affect associations survived correction for multiple comparisons either. This pattern is consistent with evidence that emotional register rather than topical content is a primary driver of language-physiology associations~\citep{gross2002, kiecoltglaser2002}, though it should be treated as hypothesis-generating.} This has direct practical implications: systems that route users by concern category will discard the within-person tonal variation that carries physiological relevance.

General pretrained embeddings \rev{performed as well as or better than} domain-adapted embeddings for most outcomes, consistent with prior work showing domain-specific training does not always transfer when target text differs in length and register from the pretraining corpus~\citep{gururangan2020}. \rev{The two embedding approaches were comparable on a minority of outcomes, but domain adaptation did not show a clear advantage for any outcome.}

A substantial proportion of health-relevant linguistic signal is recoverable only through distributional representations. That 32\% of significant RoBERTa dimensions and 17\% of MentalRoBERTa dimensions were not recoverable by any named SEANCE construct points to pragmatic and stylistic features---hedging, emotional minimization, brevity as a stance---that dictionary approaches will systematically miss, consistent with known limitations of lexicon-based NLP for capturing contextual language~\citep{pennebaker2003, tausczik2010}.

\subsection{Methodological Considerations}
\label{sec:disc-methods}

The language block $\Delta R^2$ values in this study are modest in absolute terms (range: $0.0005$--$0.0370$) across SEANCE and both embedding models; Table \ref{tab:variance}, reflecting three structural features of the design. High between-person ICC values (0.53--0.91) leave limited within-person variance for any weekly measure to explain. The concern text is among the most minimal inputs possible: a median of three words, collected bimonthly, with no effort to elicit health-relevant language. The within-person modeling approach deliberately separates stable between-person differences from week-to-week changes, producing smaller effect sizes than between-person designs would. Future work using longer prompts, more frequent sampling, or active elicitation of affective content would likely produce larger effects.

Weeks characterized by emotional exhaustion language (MentalRoBERTa PC7) were associated with poorer sleep efficiency and lower RMSSD simultaneously, while weeks characterized by elaborated relational distress (MentalRoBERTa PC31 at the negative pole) were associated with worse outcomes across multiple domains. One interpretation is that emotional exhaustion reflects chronic unresolved distress that disrupts sleep consolidation and autonomic recovery, while relational distress reflects active emotional processing with different downstream physiological consequences, consistent with work distinguishing reactive sadness from chronic emotional dysregulation in sleep research~\citep{harvey2002, nolenhoeksema2008}. These interpretations are speculative and require direct testing.

\subsection{Limitations}
\label{sec:disc-limitations}

Several limitations constrain the claims that can be drawn from these findings. The sample was drawn from a single public research university (University of Vermont) and was predominantly White (\rev{86.9\%}) and female (\rev{64.5\%}), limiting generalizability to other demographic groups and institutional contexts. The concern text prompt was administered bimonthly rather than daily, limiting sensitivity to rapid affective fluctuations and precluding lagged analyses that would clarify the temporal direction of language-outcome associations. The within-person design cannot establish causation; physiological states may influence linguistic expression rather than, or in addition to, the reverse. \rev{The random-intercept-only specification (Section~\ref{sec:mixedeffects}) assumes that the association between a given linguistic feature and a wearable outcome is uniform across participants, with only baseline levels varying by person; if these associations instead vary substantially across individuals, the reported average effects may mask meaningful heterogeneity. Random-slope models would test this directly but were not feasible given the sparse within-person data structure here (median 8 waves per participant) and remain a priority for future work with denser sampling.} RMSSD's high ICC (0.91) severely constrains available within-person variance for language associations, and the near-zero language block $\Delta R^2$ for this outcome under RoBERTa reflects a structural limitation of the outcome rather than a failure of the linguistic features. Embedding dimension interpretation relied on triangulation with SEANCE features, domain scores, and representative texts rather than independent human coding, and direct validation remains a priority for future work. The concern prompt used in this study introduces a negatively valenced framing that may systematically shape the affective content of responses. \rev{By asking what concerns participants most, the prompt selectively surfaces 
worry, stress, and apprehension, and the associations observed in this study therefore reflect the physiological correlates of negatively valenced affective expression rather than affective language in general. Whether positively framed prompts yield complementary or distinct physiological associations is an open empirical question. Positive experiences, such as academic achievements, social connection, or moments of recovery, may produce affective language that associates with better sleep or higher physical activity in ways that a concern prompt cannot capture. A balanced prompting strategy, pairing a concern prompt with a positive counterpart in the same assessment wave, would allow direct comparison of the physiological signal recoverable from each valence and provide a more complete picture of how affective language complements passive sensing. The present study establishes what a negatively framed prompt can recover; 
future work should establish whether and how much a positively framed prompt adds}

\subsection{Design Implications for Wearable Sensing in Digital Health}
\label{sec:disc-implications}

\paragraph{Ultra-brief affective prompts as a complement to passive sensing.}
Wearable devices have become increasingly sophisticated at capturing physiological and behavioral signals, yet they remain psychologically silent---they can detect that a student's sleep efficiency declined this week but cannot tell you why. Commercial wearable platforms have begun to address this gap through structured user input: WHOOP's daily journal prompts users to log behaviors such as alcohol consumption, caffeine intake, and stress level; Oura's readiness feature incorporates user-reported sleep quality; and Garmin's stress tracking combines heart rate variability with activity data to generate a composite stress score~\citep{stanton2023}. When Oura detects a particularly good or bad night of sleep, the app prompts users to tag it from a predefined list or enter a custom label, a design that implicitly acknowledges the value of user-generated language for interpreting physiological data. These findings suggest that open-ended text goes further: even \rev{four-word median} responses capture affective signal beyond what topically structured inputs can recover. Topic classification specifically, not user-generated input in general, showed no reliable associations with physiological outcomes in the present results\rev{, while affective language showed more associations, though these were not confirmed by correction for multiple comparisons}. HCI research on personal tracking has similarly shown that users interpret wearable data in subjective and contextually embedded ways that structured interfaces often fail to capture~\citep{rooksby2014}. Taken together, these observations point toward a design direction: pairing wearable sensing with ultra-brief, open-ended affective text as a low-burden mechanism for capturing psychological context that passive sensors cannot recover.

Ultra-brief open-ended prompts can be integrated into wearable study protocols at minimal burden. Even three-word responses carry detectable affective signal above and beyond objective physiological trends, suggesting that the threshold for useful text input is far lower than prior work relying on longer diary entries or multi-item EMA batteries has assumed~\citep{heron2017, wen2017}. A single bimonthly question adds negligible burden to a wearable protocol while providing psychological context that the wearable cannot. This positions ultra-brief affective text as a practical complement to passive sensing rather than a competing modality, consistent with calls for low-intensity approaches that scale across large and diverse populations~\citep{schleider2017, schleider2020}.

\paragraph{Prompt design: prioritize affective register over topic.}
Affective feature extraction \rev{may be a more promising direction than} topic classification as the primary text processing step\rev{, though this should be treated as a preliminary finding rather than an established result}. The zero-shot classification results demonstrate that what students are
worried about showed only one association with a physiological outcome surviving correction; affective dimensions showed more associations, though most were uncorrected. Systems that route users by concern category \rev{may discard} within-person tonal variation \rev{that these results suggest carries} physiological relevance. Prior work on language-based digital phenotyping has similarly emphasized affective and stylistic features over topical classification as more reliable markers of psychological state~\citep{pennebaker2003, tausczik2010}, and the present results \rev{offer preliminary, hypothesis-generating support for extending} this principle to the specific context of brief longitudinal text paired with wearable data.

\paragraph{\rev{Prompt design: what makes a brief prompt effective.}}
\rev{The findings also offer preliminary guidance on prompt design. The concern prompt used in this study---``\textit{Of all the things that have happened in the last 2 weeks, what concerns you the most?}''---is negatively valenced by construction, and the results show that even within this constrained framing, affective variation in responses associates with physiological outcomes. This suggests that the key design property of an effective ultra-brief affective prompt is not the specific topic it elicits but its capacity to surface emotional register: prompts that invite free expression of psychological state, whether framed around concerns, highlights, or anticipated stressors, are likely 
to yield affectively variable responses that carry physiological signal. Future work should examine whether positively framed prompts (e.g., ``\textit{What are 
you looking forward to this week?}'') yield complementary or distinct associations, and whether prompts explicitly targeting emotional state (e.g., ``\textit{How are you feeling right now?}'') outperform topic-anchored prompts in recovering physiologically relevant signal.}

\paragraph{\rev{Prompt frequency and the burden--information gain trade-off.}}
\rev{On the question of prompt frequency, the present results establish that a bimonthly prompt, one of the lowest conceivable sampling rates, carries detectable within-person signal, suggesting that the information gain threshold for brief affective text is lower than prior work relying on daily or weekly EMA has assumed~\cite{heron2017, wen2017}. The appropriate frequency will depend on the application: passive monitoring protocols aiming to detect gradual drift in wellbeing may find bimonthly prompts sufficient, while just-in-time adaptive intervention systems~\cite{nahumshani2018} requiring rapid detection of acute distress would benefit from higher-frequency sampling. The burden--information gain trade-off is not fixed: because the signal lies in emotional tone rather than topical elaboration, prompts can remain ultra-brief even at higher frequencies without requiring greater participant effort. A single open-ended question added to an existing weekly check-in, for instance, would increase sampling frequency sixfold relative to the present study while adding negligible burden.}

\paragraph{NLP method selection.}
For NLP method selection, general pretrained embeddings are a reasonable default for short-text passive sensing in student populations. \rev{Deploying a model such as RoBERTa-base requires no domain-specific training data and performed as well as or better than the domain-adapted model across most outcomes in this study, including autonomic outcomes such as RMSSD, lowering the barrier for researchers and practitioners seeking to integrate text-based features into wearable sensing pipelines. Domain adaptation did not show a clear advantage for any outcome, though
its performance was comparable to general embeddings on a minority of outcomes; whether domain-specific pretraining offers benefits for other populations or text registers remains an open question.}

\paragraph{\rev{Implementation hurdles.}}
\rev{Realizing this design pattern in practice requires addressing several concrete hurdles. Prompt timing relative to physiological measurement windows matters: a concern expressed at the end of a two-week window may not align temporally with the wearable data it is meant to contextualize, and future systems should explore real-time or event-contingent prompting~\cite{nahumshani2018}. Privacy is a meaningful barrier: open-ended affective text is more sensitive than structured behavioral logs, and systems that pair such text with physiological data require transparent data governance and participant control over what is stored and processed. Translating affective signals into actionable outputs is a third challenge: knowing that a student's concern text this week reflects emotional exhaustion is only useful if it can trigger a timely and appropriate response, whether an automated nudge, a clinician alert, or a self-reflection prompt. These implementation challenges represent a concrete agenda for future HCI and ubiquitous computing research, and the present findings provide the empirical foundation, establishing that the signal exists and where it lies, on which that agenda can build.}
\section{Conclusion}
\label{sec:conclusion}

Wearables measure biology precisely but cannot recover the psychological context shaping what they measure. This study demonstrates that ultra-brief, naturalistic concern text---responses with a median length of three words---carries detectable affective signal about concurrent sleep and physical activity outcomes within individuals across a full academic year, above and beyond semester-level trends. The signal lies in emotional tone, not topical content: zero-shot classification of what students worried about produced only one association surviving multiple-comparison correction, while affective dimensions across all three NLP methods were consistently associated with wearable outcomes. Embedding-based representations substantially outperformed dictionary methods, and a substantial proportion of health-relevant linguistic signal was recoverable only through distributional representations. These findings establish a lower bound on what passive affective text sensing can achieve and position ultra-brief open-ended prompts as a practical, low-burden complement to wearable sensing---one that provides the psychological context that passive sensors cannot, at minimal cost to participants or researchers. Future work should examine generalizability across populations, institutions, and cultural contexts, and explore how affective signals derived from brief text can be integrated into context-aware systems that support timely outreach to students who may benefit.



\begin{acks}
The authors acknowledge financial support from the MassMutual Center of Excellence in Complex Systems and the National Science Foundation via awards \#2242829 and \#2419830. Claude (Anthropic) was used to assist with analysis code generation, figure production, and manuscript editing and revision. All analytical decisions, interpretations, and conclusions are the authors' own.
\end{acks}

\bibliographystyle{ACM-Reference-Format}
\bibliography{citations}

\appendix
\clearpage
\section{Demographics}
\label{app:demographics}
Demographic characteristics for the analytic sample are reported in Table~\ref{tab:demographics}. Baseline survey data were available for 451 of 458 participants (98.5\%). Seven participants could not be matched to the baseline file.
\begin{table}[H]
\centering
\caption{Demographic Characteristics of the Analytic Sample}
\label{tab:demographics}
\begin{tabular}{lrr}
\hline
Characteristic & $n$ & \% \\
\hline
\multicolumn{3}{l}{\textit{Sample}} \\
Total wearable participants & 458 & --- \\
Matched to baseline demographics & 451 & --- \\
Age, $M$ ($SD$) & 18.6 (0.6) & --- \\
Age range & 18--21 years & --- \\
\multicolumn{3}{l}{\textit{Gender identity (multi-select)}} \\
\quad Woman & 291 & 64.5 \\
\quad Man & 114 & 25.3 \\
\quad Genderqueer/non-binary & 35 & 7.8 \\
\quad Transgender & 15 & 3.3 \\
\quad Agender & 5 & 1.1 \\
\quad Other & 6 & 1.3 \\
\multicolumn{3}{l}{\textit{Race/ethnicity (multi-select)}} \\
\quad White & 392 & 86.9 \\
\quad European & 76 & 16.9 \\
\quad Hispanic/Latino & 32 & 7.1 \\
\quad East Asian & 19 & 4.2 \\
\quad Middle Eastern & 13 & 2.9 \\
\quad Southeast Asian & 10 & 2.2 \\
\quad African American & 8 & 1.8 \\
\quad Native American & 4 & 0.9 \\
\quad African Caribbean & 3 & 0.7 \\
\quad Other & 14 & 3.1 \\
\multicolumn{3}{l}{\textit{Sexual orientation (multi-select)}} \\
\quad Straight/heterosexual & 231 & 51.2 \\
\quad Bisexual & 108 & 23.9 \\
\quad Queer & 43 & 9.5 \\
\quad Gay/lesbian & 33 & 7.3 \\
\quad Questioning & 33 & 7.3 \\
\quad Pansexual & 20 & 4.4 \\
\quad Asexual & 11 & 2.4 \\
\quad Other & 6 & 1.3 \\
\hline
\end{tabular}
\begin{tablenotes}
\small
\item \textit{Note.} $N = 458$ total wearable participants; demographic data available for 451 matched to baseline. Seven participants could not be matched to the baseline file. Gender identity, race/ethnicity, and sexual orientation were multi-select; percentages are calculated out of matched $N = 451$ and may sum to more than 100\%. Age reported as mean ($SD$).
\end{tablenotes}
\end{table}
\FloatBarrier


\section{SEANCE Dominance Analysis}
\label{app:dominance}
Table~\ref{tab:dominance} reports the full dominance analysis results for SEANCE features, showing the average marginal $R^2$ contribution of each predictor across all possible model subsets for outcomes with at least one significant SEANCE association ($p < .05$). Semester timing (\texttt{Week\_z}) is included as a competitor to quantify its dominance relative to linguistic features.
\begin{longtable}{llrr}
\caption{SEANCE Dominance Analysis: Average Marginal $R^2$ Contributions Across All Predictor Subsets}
\label{tab:dominance} \\
\toprule
Outcome & Predictor & Avg $\Delta R^2$ & \% of explained $R^2$ \\
\midrule
\endfirsthead
\multicolumn{4}{l}{\textit{Table~\ref{tab:dominance} continued}} \\
\toprule
Outcome & Predictor & Avg $\Delta R^2$ & \% of explained $R^2$ \\
\midrule
\endhead
\midrule
\multicolumn{4}{r}{\textit{Continued on next page}} \\
\endfoot
\bottomrule
\endlastfoot

\multicolumn{4}{l}{\textit{RMSSD (ms)}} \\
\rev{ & Week (semester timing) & 0.000911 & $+55.6\%$ \\
 & \texttt{Academ\_GI} (academic language) & 0.000240 & $+14.6\%$ \\
 & \texttt{Valence} (ANEW) & 0.000158 & $+9.6\%$ \\
 & \texttt{Sadness\_EmoLex} (sadness) & 0.000149 & $+9.1\%$ \\
 & \texttt{Negative\_EmoLex} (negative affect) & 0.000064 & $+3.9\%$ \\
 & \texttt{Strong\_GI} (forceful language) & 0.000057 & $+3.5\%$ \\
 & \texttt{Fear\_EmoLex} (fear language) & 0.000042 & $+2.6\%$ \\
 & \texttt{Work\_GI} (task-focused language) & 0.000019 & $+1.1\%$ \\}
\midrule

\multicolumn{4}{l}{\textit{Sleep Efficiency (\%)}} \\
\rev{ & Week (semester timing) & 0.002314 & $+56.4\%$ \\
 & \texttt{Work\_GI} (task-focused language) & 0.000526 & $+12.8\%$ \\
 & \texttt{Academ\_GI} (academic language) & 0.000404 & $+9.8\%$ \\
 & \texttt{Fear\_EmoLex} (fear language) & 0.000315 & $+7.7\%$ \\
 & \texttt{Strong\_GI} (forceful language) & 0.000241 & $+5.9\%$ \\
 & \texttt{Valence} (ANEW) & 0.000168 & $+4.1\%$ \\
 & \texttt{Negative\_EmoLex} (negative affect) & 0.000073 & $+1.8\%$ \\
 & \texttt{Sadness\_EmoLex} (sadness) & 0.000065 & $+1.6\%$ \\}
\midrule

\multicolumn{4}{l}{\textit{Sleep Onset Latency (min)}} \\
\rev{ & Week (semester timing) & 0.001491 & $+38.8\%$ \\
 & \texttt{Academ\_GI} (academic language) & 0.001392 & $+36.2\%$ \\
 & \texttt{Valence} (ANEW) & 0.000358 & $+9.3\%$ \\
 & \texttt{Strong\_GI} (forceful language) & 0.000241 & $+6.3\%$ \\
 & \texttt{Work\_GI} (task-focused language) & 0.000183 & $+4.8\%$ \\
 & \texttt{Fear\_EmoLex} (fear language) & 0.000130 & $+3.4\%$ \\
 & \texttt{Sadness\_EmoLex} (sadness) & 0.000030 & $+0.8\%$ \\
 & \texttt{Negative\_EmoLex} (negative affect) & 0.000017 & $+0.4\%$ \\}
\midrule

\multicolumn{4}{l}{\textit{Deep Sleep (hrs)}} \\
\rev{ & \texttt{Sadness\_EmoLex} (sadness) & 0.000497 & $+43.5\%$ \\
& \texttt{Strong\_GI} (forceful language) & 0.000164 & $+14.4\%$ \\
& \texttt{Valence} (ANEW) & 0.000138 & $+12.1\%$ \\
& \texttt{Fear\_EmoLex} (fear language) & 0.000130 & $+11.4\%$ \\
& \texttt{Negative\_EmoLex} (negative affect) & 0.000101 & $+8.8\%$ \\
& Week (semester timing) & 0.000068 & $+6.0\%$ \\
& \texttt{Work\_GI} (task-focused language) & 0.000037 & $+3.2\%$ \\
& \texttt{Academ\_GI} (academic language) & 0.000007 & $+0.6\%$ \\}
\midrule

\multicolumn{4}{l}{\textit{Steps/day}} \\
\rev{ & Week (semester timing) & 0.003193 & $+54.0\%$ \\
 & \texttt{Strong\_GI} (forceful language) & 0.002120 & $+35.8\%$ \\
 & \texttt{Work\_GI} (task-focused language) & 0.000416 & $+7.0\%$ \\
 & \texttt{Academ\_GI} (academic language) & 0.000146 & $+2.5\%$ \\
 & \texttt{Valence} (ANEW) & 0.000022 & $+0.4\%$ \\
 & \texttt{Negative\_EmoLex} (negative affect) & 0.000011 & $+0.2\%$ \\
 & \texttt{Sadness\_EmoLex} (sadness) & 0.000005 & $+0.1\%$ \\
 & \texttt{Fear\_EmoLex} (fear language) & 0.000001 & $+0.0\%$ \\}
\midrule

\multicolumn{4}{l}{\textit{MET-min Medium}} \\
\rev{ & \texttt{Strong\_GI} (forceful language) & 0.002263 & $+60.6\%$ \\
 & Week (semester timing) & 0.000563 & $+15.1\%$ \\
 & \texttt{Work\_GI} (task-focused language) & 0.000472 & $+12.7\%$ \\
 & \texttt{Academ\_GI} (academic language) & 0.000178 & $+4.8\%$ \\
 & \texttt{Negative\_EmoLex} (negative affect) & 0.000114 & $+3.1\%$ \\
 & \texttt{Fear\_EmoLex} (fear language) & 0.000071 & $+1.9\%$ \\
 & \texttt{Sadness\_EmoLex} (sadness) & 0.000049 & $+1.3\%$ \\
 & \texttt{Valence} (ANEW) & 0.000021 & $+0.6\%$ \\}
\midrule

\multicolumn{4}{l}{\textit{MET-min High}} \\
\rev{ & Week (semester timing) & 0.000892 & $+45.9\%$ \\
 & \texttt{Valence} (ANEW) & 0.000269 & $+13.8\%$ \\
 & \texttt{Work\_GI} (task-focused language) & 0.000257 & $+13.2\%$ \\
 & \texttt{Strong\_GI} (forceful language) & 0.000196 & $+10.1\%$ \\
 & \texttt{Negative\_EmoLex} (negative affect) & 0.000094 & $+4.9\%$ \\
 & \texttt{Academ\_GI} (academic language) & 0.000093 & $+4.8\%$ \\
 & \texttt{Sadness\_EmoLex} (sadness) & 0.000079 & $+4.1\%$ \\
 & \texttt{Fear\_EmoLex} (fear language) & 0.000063 & $+3.3\%$ \\}

\end{longtable}

\noindent\small\textit{Note.} Dominance analysis \citep{azen2003} run on concern-present rows only ($N = 2{,}982$--$2{,}966$). Predictors competed: \texttt{Week\_z} (semester timing) and seven SEANCE linguistic features, all $z$-scored. Covariates held constant: \texttt{nwords\_z}. Avg $\Delta R^2$ = average marginal $R^2$ contribution across all $2^8 = 256$ predictor subsets. \% of explained $R^2$ = avg $\Delta R^2$ as a percentage of total explained $R^2$ (sum of positive avg $\Delta R^2$ values). Outcomes shown are those with at least one significant SEANCE association ($p < .05$). Marginal $R^2$ computed using the \citet{nakagawa2013} method.\normalsize

\section{Embedding Model Dominance Analysis}
\label{app:embedding-dominance}

Tables~\ref{tab:roberta-dominance} and~\ref{tab:mentalroberta-dominance} report dominance analysis results for RoBERTa-base and MentalRoBERTa respectively, showing the average marginal $R^2$ contribution of each significant PC across all possible model subsets for each outcome. Semester timing (\texttt{Week\_z}) is included as a competitor.

\begin{longtable}{llrr}
\caption{RoBERTa-Base Dominance Analysis: Average Marginal $R^2$ Contributions}
\label{tab:roberta-dominance} \\
\toprule
Outcome & Predictor & Avg $\Delta R^2$ & \% of explained $R^2$ \\
\midrule
\endfirsthead
\multicolumn{4}{l}{\textit{Table~\ref{tab:roberta-dominance} continued}} \\
\toprule
Outcome & Predictor & Avg $\Delta R^2$ & \% of explained $R^2$ \\
\midrule
\endhead
\midrule
\multicolumn{4}{r}{\textit{Continued on next page}} \\
\endfoot
\bottomrule
\endlastfoot

\multicolumn{4}{l}{\textit{Sleep duration (hrs)}} \\
 & PC114 & 0.001429 & $+15.9\%$ \\
 & PC117 & 0.001282 & $+14.2\%$ \\
 & PC9   & 0.001250 & $+13.9\%$ \\
 & PC119 & 0.001225 & $+13.6\%$ \\
 & PC62  & 0.001113 & $+12.4\%$ \\
 & PC83  & 0.001014 & $+11.2\%$ \\
 & PC41  & 0.000944 & $+10.5\%$ \\
 & PC16  & 0.000755 & $+8.4\%$ \\
\midrule

\multicolumn{4}{l}{\textit{Sleep efficiency (\%)}} \\
 & PC34  & 0.001278 & $+16.8\%$ \\
 & PC116 & 0.001038 & $+13.7\%$ \\
 & PC86  & 0.000820 & $+10.8\%$ \\
 & PC30  & 0.000750 & $+9.9\%$ \\
 & PC39  & 0.000743 & $+9.8\%$ \\
 & PC25  & 0.000715 & $+9.4\%$ \\
 & PC48  & 0.000624 & $+8.2\%$ \\
 & PC43  & 0.000565 & $+7.4\%$ \\
 & PC109 & 0.000559 & $+7.4\%$ \\
 & PC46  & 0.000516 & $+6.8\%$ \\
\midrule

\multicolumn{4}{l}{\textit{REM sleep (hrs)}} \\
 & PC63  & 0.001601 & $+26.4\%$ \\
 & PC7   & 0.001117 & $+18.4\%$ \\
 & PC9   & 0.000933 & $+15.4\%$ \\
 & PC117 & 0.000896 & $+14.8\%$ \\
 & PC119 & 0.000691 & $+11.4\%$ \\
 & PC38  & 0.000490 & $+8.1\%$ \\
 & PC41  & 0.000342 & $+5.6\%$ \\
\midrule

\multicolumn{4}{l}{\textit{Deep sleep (hrs)}} \\
 & PC56  & 0.000855 & $+16.3\%$ \\
 & PC77  & 0.000815 & $+15.5\%$ \\
 & PC49  & 0.000644 & $+12.2\%$ \\
 & PC124 & 0.000611 & $+11.6\%$ \\
 & PC120 & 0.000570 & $+10.8\%$ \\
 & PC114 & 0.000480 & $+9.1\%$ \\
 & PC63  & 0.000476 & $+9.0\%$ \\
 & PC47  & 0.000427 & $+8.1\%$ \\
 & PC97  & 0.000381 & $+7.2\%$ \\
\midrule

\multicolumn{4}{l}{\textit{RMSSD (ms)}} \\
 & PC45  & 0.000308 & $+16.4\%$ \\
 & PC41  & 0.000298 & $+15.9\%$ \\
 & PC29  & 0.000297 & $+15.8\%$ \\
 & PC7   & 0.000190 & $+10.1\%$ \\
 & PC115 & 0.000188 & $+10.0\%$ \\
 & PC92  & 0.000179 & $+9.5\%$ \\
 & PC32  & 0.000174 & $+9.3\%$ \\
 & PC112 & 0.000144 & $+7.6\%$ \\
 & PC124 & 0.000103 & $+5.5\%$ \\
\midrule

\multicolumn{4}{l}{\textit{Sleep onset latency (min)}} \\
 & PC34  & 0.001404 & $+16.9\%$ \\
 & PC5   & 0.001368 & $+16.5\%$ \\
 & PC90  & 0.001057 & $+12.8\%$ \\
 & PC39  & 0.000939 & $+11.3\%$ \\
 & PC58  & 0.000799 & $+9.6\%$ \\
 & PC57  & 0.000767 & $+9.3\%$ \\
 & PC94  & 0.000732 & $+8.8\%$ \\
 & PC110 & 0.000676 & $+8.2\%$ \\
 & PC87  & 0.000547 & $+6.6\%$ \\
\midrule

\multicolumn{4}{l}{\textit{Steps/day}} \\
 & PC30  & 0.001785 & $+22.8\%$ \\
 & PC117 & 0.000868 & $+11.1\%$ \\
 & PC52  & 0.000783 & $+10.0\%$ \\
 & PC82  & 0.000753 & $+9.6\%$ \\
 & PC54  & 0.000676 & $+8.6\%$ \\
 & PC74  & 0.000657 & $+8.4\%$ \\
 & PC75  & 0.000656 & $+8.4\%$ \\
 & PC105 & 0.000584 & $+7.5\%$ \\
 & PC97  & 0.000546 & $+7.0\%$ \\
 & PC109 & 0.000534 & $+6.8\%$ \\
\midrule

\multicolumn{4}{l}{\textit{MET-min medium}} \\
 & PC13  & 0.003503 & $+22.5\%$ \\
 & PC6   & 0.002973 & $+19.1\%$ \\
 & PC52  & 0.001622 & $+10.4\%$ \\
 & PC28  & 0.001396 & $+9.0\%$ \\
 & PC30  & 0.001190 & $+7.6\%$ \\
 & PC26  & 0.001147 & $+7.4\%$ \\
 & PC94  & 0.001058 & $+6.8\%$ \\
 & PC117 & 0.000933 & $+6.0\%$ \\
 & PC74  & 0.000909 & $+5.8\%$ \\
 & PC92  & 0.000832 & $+5.3\%$ \\
\midrule

\multicolumn{4}{l}{\textit{MET-min high}} \\
 & PC110 & 0.001104 & $+25.1\%$ \\
 & PC51  & 0.000821 & $+18.6\%$ \\
 & PC27  & 0.000678 & $+15.4\%$ \\
 & PC40  & 0.000669 & $+15.2\%$ \\
 & PC30  & 0.000636 & $+14.4\%$ \\
 & PC73  & 0.000497 & $+11.3\%$ \\

\end{longtable}

\noindent\small\textit{Note.} Dominance analysis run on concern-present rows only ($N = 2{,}982$ sleep outcomes; $N = 2{,}966$ activity outcomes). Predictors are the principal components significant ($p<.05$) in the joint model for that outcome, capped at the ten with the largest absolute regression coefficients; the base model contains \texttt{Week\_z} and \texttt{nwords\_z} and is estimated by maximum likelihood with a participant random intercept. Avg $\Delta R^2$ = average increment in fixed-effects marginal $R^2$ across all possible predictor subsets. \% of explained $R^2$ = avg $\Delta R^2$ as a percentage of the sum of positive avg $\Delta R^2$ values in that outcome, so percentages sum to 100\% within an outcome. Marginal $R^2$ is the variance of the fixed-effects linear predictor divided by total variance \citep{nakagawa2013}.\normalsize

\begin{longtable}{llrr}
\caption{MentalRoBERTa Dominance Analysis: Average Marginal $R^2$ Contributions}
\label{tab:mentalroberta-dominance} \\
\toprule
Outcome & Predictor & Avg $\Delta R^2$ & \% of explained $R^2$ \\
\midrule
\endfirsthead
\multicolumn{4}{l}{\textit{Table~\ref{tab:mentalroberta-dominance} continued}} \\
\toprule
Outcome & Predictor & Avg $\Delta R^2$ & \% of explained $R^2$ \\
\midrule
\endhead
\midrule
\multicolumn{4}{r}{\textit{Continued on next page}} \\
\endfoot
\bottomrule
\endlastfoot

\multicolumn{4}{l}{\textit{Sleep duration (hrs)}} \\
 & PC54  & 0.001395 & $+30.9\%$ \\
 & PC57  & 0.001075 & $+23.8\%$ \\
 & PC97  & 0.001037 & $+23.0\%$ \\
 & PC35  & 0.001010 & $+22.4\%$ \\
\midrule

\multicolumn{4}{l}{\textit{Sleep efficiency (\%)}} \\
 & PC20  & 0.002051 & $+17.2\%$ \\
 & PC10  & 0.001700 & $+14.3\%$ \\
 & PC30  & 0.001353 & $+11.3\%$ \\
 & PC31  & 0.001290 & $+10.8\%$ \\
 & PC18  & 0.001124 & $+9.4\%$ \\
 & PC7   & 0.001123 & $+9.4\%$ \\
 & PC22  & 0.000882 & $+7.4\%$ \\
 & PC26  & 0.000874 & $+7.3\%$ \\
 & PC75  & 0.000801 & $+6.7\%$ \\
 & PC87  & 0.000725 & $+6.1\%$ \\
\midrule

\multicolumn{4}{l}{\textit{REM sleep (hrs)}} \\
 & PC22  & 0.000952 & $+37.2\%$ \\
 & PC31  & 0.000681 & $+26.6\%$ \\
 & PC97  & 0.000548 & $+21.4\%$ \\
 & PC52  & 0.000381 & $+14.9\%$ \\
\midrule

\multicolumn{4}{l}{\textit{Deep sleep (hrs)}} \\
 & PC30  & 0.001007 & $+26.3\%$ \\
 & PC46  & 0.000809 & $+21.1\%$ \\
 & PC66  & 0.000749 & $+19.5\%$ \\
 & PC64  & 0.000458 & $+11.9\%$ \\
 & PC82  & 0.000451 & $+11.8\%$ \\
 & PC41  & 0.000358 & $+9.3\%$ \\
\midrule

\multicolumn{4}{l}{\textit{RMSSD (ms)}} \\
 & PC31  & 0.000366 & $+18.0\%$ \\
 & PC7   & 0.000334 & $+16.4\%$ \\
 & PC71  & 0.000214 & $+10.5\%$ \\
 & PC47  & 0.000191 & $+9.4\%$ \\
 & PC29  & 0.000191 & $+9.4\%$ \\
 & PC85  & 0.000181 & $+8.9\%$ \\
 & PC27  & 0.000156 & $+7.7\%$ \\
 & PC39  & 0.000152 & $+7.5\%$ \\
 & PC45  & 0.000135 & $+6.7\%$ \\
 & PC60  & 0.000115 & $+5.6\%$ \\
\midrule

\multicolumn{4}{l}{\textit{Sleep onset latency (min)}} \\
 & PC24  & 0.001139 & $+18.8\%$ \\
 & PC38  & 0.000929 & $+15.3\%$ \\
 & PC72  & 0.000909 & $+15.0\%$ \\
 & PC54  & 0.000851 & $+14.0\%$ \\
 & PC71  & 0.000824 & $+13.6\%$ \\
 & PC20  & 0.000756 & $+12.4\%$ \\
 & PC91  & 0.000667 & $+11.0\%$ \\
\midrule

\multicolumn{4}{l}{\textit{Steps/day}} \\
 & PC34  & 0.001256 & $+16.1\%$ \\
 & PC13  & 0.001211 & $+15.6\%$ \\
 & PC73  & 0.000883 & $+11.3\%$ \\
 & PC47  & 0.000880 & $+11.3\%$ \\
 & PC44  & 0.000675 & $+8.7\%$ \\
 & PC70  & 0.000665 & $+8.5\%$ \\
 & PC29  & 0.000601 & $+7.7\%$ \\
 & PC52  & 0.000582 & $+7.5\%$ \\
 & PC59  & 0.000536 & $+6.9\%$ \\
 & PC55  & 0.000501 & $+6.4\%$ \\
\midrule

\multicolumn{4}{l}{\textit{MET-min medium}} \\
 & PC13  & 0.001976 & $+15.3\%$ \\
 & PC5   & 0.001568 & $+12.2\%$ \\
 & PC15  & 0.001472 & $+11.4\%$ \\
 & PC73  & 0.001432 & $+11.1\%$ \\
 & PC47  & 0.001416 & $+11.0\%$ \\
 & PC6   & 0.001402 & $+10.9\%$ \\
 & PC11  & 0.001175 & $+9.1\%$ \\
 & PC34  & 0.000971 & $+7.5\%$ \\
 & PC52  & 0.000820 & $+6.4\%$ \\
 & PC17  & 0.000644 & $+5.0\%$ \\
\midrule

\multicolumn{4}{l}{\textit{MET-min high}} \\
 & PC29  & 0.000921 & $+20.6\%$ \\
 & PC81  & 0.000562 & $+12.6\%$ \\
 & PC31  & 0.000559 & $+12.5\%$ \\
 & PC70  & 0.000526 & $+11.8\%$ \\
 & PC59  & 0.000525 & $+11.7\%$ \\
 & PC34  & 0.000479 & $+10.7\%$ \\
 & PC69  & 0.000467 & $+10.5\%$ \\
 & PC85  & 0.000426 & $+9.5\%$ \\

\end{longtable}

\noindent\small\textit{Note.} Concern-present rows only ($N = 2{,}982$ sleep outcomes; $N = 2{,}966$ activity outcomes). Structure and abbreviations follow Table~\ref{tab:roberta-dominance}.\normalsize

\section{RoBERTa-Base: Full Coefficient Table}
\label{app:roberta}

Table~\ref{tab:roberta-full} reports all significant RoBERTa-base PCA component associations with wearable outcomes ($p < .05$, uncorrected). Components are sorted within each outcome block by direction (positive then negative) then $p$-value. The Note column indicates cross-outcome associations for the same PC.

\begin{longtable}{lrrrll}
\caption{RoBERTa-Base: Significant PC Associations with Wearable Outcomes ($p < .05$)}
\label{tab:roberta-full} \\
\toprule
\textbf{PC} & \textbf{$\beta$} & \textbf{SE} & \textbf{$p$} & \textbf{95\% CI} & \textbf{Note} \\
\midrule
\endfirsthead
\multicolumn{6}{l}{\textit{Table~\ref{tab:roberta-full} continued}} \\
\toprule
\textbf{PC} & \textbf{$\beta$} & \textbf{SE} & \textbf{$p$} & \textbf{95\% CI} & \textbf{Note} \\
\midrule
\endhead
\midrule
\multicolumn{6}{r}{\textit{Continued on next page}} \\
\endfoot
\bottomrule
\endlastfoot

\multicolumn{6}{l}{\textit{MET-min high (6 significant PCs)}} \\
PC110 & $+3.431$ & 1.084 & $= .002$ & $[+1.31, +5.56]$ & \\
PC51 & $+2.060$ & 0.729 & $= .005$ & $[+0.63, +3.49]$ & \\
PC40 & $+1.504$ & 0.614 & $= .014$ & $[+0.30, +2.71]$ & \\
PC73 & $+1.842$ & 0.844 & $= .029$ & $[+0.19, +3.50]$ & \\
PC30 & $-1.318$ & 0.548 & $= .016$ & $[-2.39, -0.24]$ & \\
PC27 & $-1.410$ & 0.549 & $= .010$ & $[-2.48, -0.33]$ & \\
\midrule

\multicolumn{6}{l}{\textit{MET-min medium (11 significant PCs)}} \\
PC13 & $+2.083$ & 0.485 & $< .001$ & $[+1.13, +3.03]$ & Also sig.: steps/day \\
PC52 & $+2.442$ & 0.855 & $= .004$ & $[+0.77, +4.12]$ & \\
PC117 & $+3.177$ & 1.313 & $= .016$ & $[+0.60, +5.75]$ & Also sig.: sleep dur., REM, steps/day \\
PC4 & $+0.812$ & 0.341 & $= .017$ & $[+0.14, +1.48]$ & Also sig.: steps/day \\
PC6 & $-1.294$ & 0.387 & $= .001$ & $[-2.05, -0.54]$ & Also sig.: steps/day \\
PC28 & $-1.622$ & 0.639 & $= .011$ & $[-2.87, -0.37]$ & Also sig.: steps/day \\
PC26 & $-1.437$ & 0.609 & $= .018$ & $[-2.63, -0.24]$ & Also sig.: steps/day \\
PC30 & $-1.629$ & 0.646 & $= .012$ & $[-2.89, -0.36]$ & Also sig.: sleep eff., steps/day, MET-high \\
PC74 & $-2.495$ & 1.019 & $= .014$ & $[-4.49, -0.50]$ & Also sig.: steps/day \\
PC92 & $-2.433$ & 1.107 & $= .028$ & $[-4.60, -0.26]$ & \\
PC94 & $-2.694$ & 1.131 & $= .017$ & $[-4.91, -0.48]$ & \\
\midrule

\multicolumn{6}{l}{\textit{Steps/day (19 significant PCs)}} \\
PC13 & $+55.071$ & 13.547 & $< .001$ & $[+28.52, +81.62]$ & Also sig.: MET-med \\
PC22 & $+45.957$ & 16.673 & $= .006$ & $[+13.28, +78.64]$ & \\
PC52 & $+63.556$ & 23.816 & $= .008$ & $[+16.88, +110.23]$ & \\
PC54 & $+59.816$ & 24.028 & $= .013$ & $[+12.72, +106.91]$ & \\
PC97 & $+66.462$ & 32.592 & $= .041$ & $[+2.58, +130.34]$ & \\
PC109 & $+72.600$ & 35.305 & $= .040$ & $[+3.40, +141.80]$ & \\
PC117 & $+96.487$ & 36.500 & $= .008$ & $[+24.95, +168.03]$ & Also sig.: sleep dur., REM, MET-med \\
PC4 & $+22.990$ & 9.574 & $= .016$ & $[+4.22, +41.76]$ & Also sig.: MET-med \\
PC1 & $-13.676$ & 6.531 & $= .036$ & $[-26.48, -0.88]$ & Academic language dimension \\
PC6 & $-32.994$ & 10.837 & $= .002$ & $[-54.23, -11.75]$ & Also sig.: MET-med \\
PC16 & $-36.903$ & 14.579 & $= .011$ & $[-65.48, -8.33]$ & \\
PC26 & $-48.347$ & 16.946 & $= .004$ & $[-81.56, -15.13]$ & Also sig.: MET-med \\
PC28 & $-52.410$ & 17.841 & $= .003$ & $[-87.38, -17.44]$ & Also sig.: MET-med \\
PC30 & $-61.671$ & 17.980 & $= .001$ & $[-96.91, -26.43]$ & Also sig.: sleep eff., MET-med, MET-high \\
PC65 & $-57.867$ & 27.491 & $= .035$ & $[-111.75, -3.99]$ & \\
PC67 & $-67.320$ & 28.345 & $= .018$ & $[-122.88, -11.77]$ & \\
PC70 & $-73.708$ & 30.154 & $= .015$ & $[-132.81, -14.61]$ & \\
PC74 & $-75.435$ & 28.384 & $= .008$ & $[-131.07, -19.80]$ & Also sig.: MET-med \\
PC105 & $-70.692$ & 33.905 & $= .037$ & $[-137.14, -4.24]$ & \\
\midrule

\multicolumn{6}{l}{\textit{Deep sleep (9 significant PCs)}} \\
PC124 & $+0.008$ & 0.003 & $= .008$ & $[+0.00, +0.01]$ & Also sig.: RMSSD \\
PC120 & $+0.007$ & 0.003 & $= .017$ & $[+0.00, +0.01]$ & \\
PC114 & $+0.006$ & 0.003 & $= .037$ & $[+0.00, +0.01]$ & Also sig.: sleep duration \\
PC49 & $-0.004$ & 0.002 & $= .016$ & $[-0.01, -0.00]$ & \\
PC56 & $-0.005$ & 0.002 & $= .005$ & $[-0.01, -0.00]$ & \\
PC63 & $-0.004$ & 0.002 & $= .027$ & $[-0.01, -0.00]$ & Also sig.: REM sleep \\
PC77 & $-0.006$ & 0.002 & $= .006$ & $[-0.01, -0.00]$ & \\
PC47 & $-0.003$ & 0.002 & $= .047$ & $[-0.01, +0.00]$ & \\
PC97 & $-0.005$ & 0.003 & $= .039$ & $[-0.01, +0.00]$ & \\
\midrule

\multicolumn{6}{l}{\textit{Sleep efficiency (12 significant PCs)}} \\
PC7 & $+0.024$ & 0.012 & $= .045$ & $[+0.00, +0.05]$ & Also sig.: REM sleep, RMSSD \\
PC39 & $+0.048$ & 0.022 & $= .030$ & $[+0.01, +0.09]$ & Also sig.: onset latency (neg.) \\
PC46 & $+0.049$ & 0.023 & $= .037$ & $[+0.00, +0.09]$ & \\
PC48 & $+0.049$ & 0.024 & $= .039$ & $[+0.00, +0.10]$ & \\
PC86 & $+0.073$ & 0.032 & $= .023$ & $[+0.01, +0.14]$ & \\
PC109 & $+0.077$ & 0.037 & $= .039$ & $[+0.00, +0.15]$ & \\
PC116 & $+0.095$ & 0.039 & $= .014$ & $[+0.02, +0.17]$ & \\
PC9 & $-0.028$ & 0.013 & $= .034$ & $[-0.05, -0.00]$ & Also sig.: sleep dur., REM sleep \\
PC25 & $-0.038$ & 0.019 & $= .040$ & $[-0.07, -0.00]$ & \\
PC30 & $-0.042$ & 0.019 & $= .027$ & $[-0.08, -0.01]$ & Also sig.: steps/day, MET-med, MET-high \\
PC34 & $-0.056$ & 0.021 & $= .008$ & $[-0.10, -0.01]$ & Also sig.: onset latency (pos.) \\
PC43 & $-0.046$ & 0.023 & $= .045$ & $[-0.09, -0.00]$ & \\
\midrule

\multicolumn{6}{l}{\textit{Sleep onset latency (9 significant PCs)}} \\
PC34 & $+0.125$ & 0.047 & $= .007$ & $[+0.03, +0.22]$ & Also sig.: sleep eff. (neg.) \\
PC57 & $+0.117$ & 0.057 & $= .039$ & $[+0.01, +0.23]$ & \\
PC58 & $+0.123$ & 0.058 & $= .034$ & $[+0.01, +0.24]$ & \\
PC90 & $+0.171$ & 0.073 & $= .019$ & $[+0.03, +0.31]$ & \\
PC94 & $+0.147$ & 0.074 & $= .046$ & $[+0.00, +0.29]$ & \\
PC5 & $-0.061$ & 0.022 & $= .006$ & $[-0.10, -0.02]$ & \\
PC39 & $-0.101$ & 0.049 & $= .037$ & $[-0.20, -0.01]$ & Also sig.: sleep eff. (pos.) \\
PC87 & $-0.141$ & 0.071 & $= .045$ & $[-0.28, -0.00]$ & \\
PC110 & $-0.172$ & 0.083 & $= .039$ & $[-0.34, -0.01]$ & Also sig.: MET-high (pos.) \\
\midrule

\multicolumn{6}{l}{\textit{REM sleep (7 significant PCs)}} \\
PC7 & $+0.002$ & 0.001 & $= .041$ & $[+0.00, +0.00]$ & Also sig.: sleep eff., RMSSD \\
PC41 & $+0.004$ & 0.002 & $= .039$ & $[+0.00, +0.01]$ & Also sig.: sleep dur., RMSSD \\
PC38 & $+0.004$ & 0.002 & $= .039$ & $[+0.00, +0.01]$ & \\
PC119 & $+0.006$ & 0.003 & $= .037$ & $[+0.00, +0.01]$ & Also sig.: sleep duration \\
PC9 & $-0.002$ & 0.001 & $= .017$ & $[-0.00, +0.00]$ & Also sig.: sleep dur., sleep eff. \\
PC63 & $-0.007$ & 0.002 & $= .002$ & $[-0.01, -0.00]$ & Also sig.: deep sleep \\
PC117 & $-0.007$ & 0.003 & $= .019$ & $[-0.01, -0.00]$ & Also sig.: sleep dur., steps/day, MET-med \\
\midrule

\multicolumn{6}{l}{\textit{RMSSD (9 significant PCs)}} \\
PC41 & $+0.315$ & 0.096 & $= .001$ & $[+0.13, +0.50]$ & Also sig.: sleep dur., REM sleep \\
PC45 & $+0.278$ & 0.100 & $= .005$ & $[+0.08, +0.47]$ & \\
PC112 & $+0.341$ & 0.165 & $= .038$ & $[+0.02, +0.66]$ & \\
PC124 & $+0.346$ & 0.174 & $= .047$ & $[+0.01, +0.69]$ & Also sig.: deep sleep \\
PC7 & $+0.112$ & 0.053 & $= .035$ & $[+0.01, +0.22]$ & Also sig.: sleep eff., REM sleep \\
PC29 & $-0.230$ & 0.082 & $= .005$ & $[-0.39, -0.07]$ & \\
PC32 & $-0.198$ & 0.085 & $= .020$ & $[-0.36, -0.03]$ & \\
PC92 & $-0.335$ & 0.140 & $= .017$ & $[-0.61, -0.06]$ & \\
PC115 & $-0.362$ & 0.165 & $= .028$ & $[-0.69, -0.04]$ & \\
\midrule

\multicolumn{6}{l}{\textit{Sleep duration (8 significant PCs)}} \\
PC41 & $+0.012$ & 0.005 & $= .010$ & $[+0.00, +0.02]$ & Also sig.: REM sleep, RMSSD \\
PC114 & $+0.021$ & 0.008 & $= .012$ & $[+0.01, +0.04]$ & Also sig.: deep sleep \\
PC119 & $+0.022$ & 0.008 & $= .011$ & $[+0.01, +0.04]$ & Also sig.: REM sleep \\
PC16 & $+0.007$ & 0.003 & $= .034$ & $[+0.00, +0.01]$ & \\
PC9 & $-0.007$ & 0.003 & $= .018$ & $[-0.01, -0.00]$ & Also sig.: sleep eff., REM sleep \\
PC62 & $-0.014$ & 0.006 & $= .015$ & $[-0.03, -0.00]$ & \\
PC83 & $-0.017$ & 0.007 & $= .014$ & $[-0.03, -0.00]$ & \\
PC117 & $-0.022$ & 0.008 & $= .010$ & $[-0.04, -0.01]$ & Also sig.: REM sleep, steps/day, MET-med \\

\end{longtable}

\noindent\textit{Note.} One joint model per outcome, fitted by REML: $outcome \sim all\ 127\ principal\ components (entered\  simultaneously) + Week_z + (1 | record_id)$, run on concern-present rows only (N = 2,982 sleep outcomes; N = 2,966 activity outcomes). PC scores are raw (not standardized), so β is the unstandardized change in the outcome per one-unit change in the PC score. SE = standard error. 95\% CI = confidence interval. p-values are uncorrected; with 127 PCs × 9 outcomes (1,143 tests), about 57 nominally significant results would be expected by chance. PCs are sorted within each outcome block by direction (positive then negative), then by p-value. The Note column lists other outcomes for which the same PC is also significant. Sleep eff. = sleep efficiency; sleep dur. = sleep duration; onset lat. = sleep onset latency; MET-med = MET-min medium; MET-high = MET-min high; pos./neg. = direction of association in the cross-referenced outcome.

\section{MentalRoBERTa: Full Coefficient Table}
\label{app:mentalroberta}

Table~\ref{tab:mentalroberta-full} reports all significant MentalRoBERTa PCA component associations with wearable outcomes ($p < .05$, uncorrected). Structure and abbreviations follow Table~\ref{tab:roberta-full}.

\begin{longtable}{lrrrll}
\caption{MentalRoBERTa: Significant PC Associations with Wearable Outcomes ($p < .05$)}
\label{tab:mentalroberta-full} \\
\toprule
\textbf{PC} & \textbf{$\beta$} & \textbf{SE} & \textbf{$p$} & \textbf{95\% CI} & \textbf{Note} \\
\midrule
\endfirsthead
\multicolumn{6}{l}{\textit{Table~\ref{tab:mentalroberta-full} continued}} \\
\toprule
\textbf{PC} & \textbf{$\beta$} & \textbf{SE} & \textbf{$p$} & \textbf{95\% CI} & \textbf{Note} \\
\midrule
\endhead
\midrule
\multicolumn{6}{r}{\textit{Continued on next page}} \\
\endfoot
\bottomrule
\endlastfoot

\multicolumn{6}{l}{\textit{MET-min high (8 significant PCs)}} \\
PC29 & $-1.575$ & 0.548 & $= .004$ & $[-2.65, -0.50]$ & Also sig.: RMSSD, steps/day \\
PC81 & $-2.068$ & 0.955 & $= .030$ & $[-3.94, -0.20]$ & \\
PC85 & $-1.931$ & 0.974 & $= .047$ & $[-3.84, -0.02]$ & Also sig.: RMSSD \\
PC31 & $+1.209$ & 0.560 & $= .031$ & $[+0.11, +2.31]$ & Also sig.: sleep eff., REM, RMSSD \\
PC34 & $+1.338$ & 0.580 & $= .021$ & $[+0.20, +2.47]$ & Also sig.: steps/day, MET-med \\
PC59 & $+1.871$ & 0.778 & $= .016$ & $[+0.35, +3.40]$ & Also sig.: steps/day \\
PC69 & $+1.996$ & 0.852 & $= .019$ & $[+0.33, +3.67]$ & \\
PC70 & $+2.003$ & 0.853 & $= .019$ & $[+0.33, +3.67]$ & Also sig.: steps/day \\
\midrule

\multicolumn{6}{l}{\textit{MET-min medium (12 significant PCs)}} \\
PC13 & $-1.655$ & 0.485 & $= .001$ & $[-2.61, -0.70]$ & Also sig.: steps/day \\
PC15 & $-1.473$ & 0.500 & $= .003$ & $[-2.45, -0.49]$ & Also sig.: steps/day \\
PC5 & $-0.828$ & 0.355 & $= .020$ & $[-1.52, -0.13]$ & \\
PC3 & $-0.716$ & 0.311 & $= .021$ & $[-1.33, -0.11]$ & Also sig.: steps/day \\
PC47 & $-2.037$ & 0.817 & $= .013$ & $[-3.64, -0.44]$ & Also sig.: RMSSD, steps/day \\
PC52 & $-1.966$ & 0.854 & $= .021$ & $[-3.64, -0.29]$ & Also sig.: REM sleep, steps/day \\
PC73 & $-2.816$ & 1.049 & $= .007$ & $[-4.87, -0.76]$ & Also sig.: steps/day \\
PC1 & $-0.388$ & 0.194 & $= .045$ & $[-0.77, -0.01]$ & Academic language dim.; also sig.: steps/day \\
PC6 & $+0.849$ & 0.362 & $= .019$ & $[+0.14, +1.56]$ & Also sig.: steps/day \\
PC11 & $+1.229$ & 0.479 & $= .010$ & $[+0.29, +2.17]$ & Also sig.: steps/day \\
PC17 & $+1.043$ & 0.524 & $= .046$ & $[+0.02, +2.07]$ & Also sig.: steps/day \\
PC34 & $+1.513$ & 0.686 & $= .027$ & $[+0.17, +2.86]$ & Also sig.: steps/day, MET-high \\
\midrule

\multicolumn{6}{l}{\textit{Steps/day (18 significant PCs)}} \\
PC34 & $+56.496$ & 19.082 & $= .003$ & $[+19.10, +93.90]$ & Also sig.: MET-med, MET-high \\
PC70 & $+78.870$ & 28.089 & $= .005$ & $[+23.82, +133.92]$ & Also sig.: MET-high \\
PC11 & $+35.072$ & 13.442 & $= .009$ & $[+8.72, +61.42]$ & Also sig.: MET-med \\
PC20 & $+39.268$ & 15.597 & $= .012$ & $[+8.70, +69.84]$ & Also sig.: sleep eff., onset lat. \\
PC59 & $+53.845$ & 25.604 & $= .035$ & $[+3.66, +104.03]$ & Also sig.: MET-high \\
PC6 & $+20.800$ & 10.134 & $= .040$ & $[+0.94, +40.66]$ & Also sig.: MET-med \\
PC17 & $+32.718$ & 14.612 & $= .025$ & $[+4.08, +61.36]$ & Also sig.: MET-med \\
PC1 & $-12.490$ & 5.531 & $= .024$ & $[-23.33, -1.65]$ & Academic language dim.; also sig.: MET-med \\
PC3 & $-26.841$ & 8.730 & $= .002$ & $[-43.95, -9.73]$ & Also sig.: MET-med \\
PC13 & $-43.863$ & 13.557 & $= .001$ & $[-70.43, -17.29]$ & Also sig.: MET-med \\
PC15 & $-43.429$ & 13.928 & $= .002$ & $[-70.73, -16.13]$ & Also sig.: MET-med \\
PC29 & $-43.495$ & 18.045 & $= .016$ & $[-78.86, -8.13]$ & Also sig.: RMSSD, MET-high \\
PC35 & $-38.134$ & 19.452 & $= .050$ & $[-76.26, -0.01]$ & Also sig.: sleep duration \\
PC44 & $-46.072$ & 21.836 & $= .035$ & $[-88.87, -3.27]$ & \\
PC47 & $-56.514$ & 22.753 & $= .013$ & $[-101.11, -11.92]$ & Also sig.: RMSSD, MET-med \\
PC52 & $-53.793$ & 23.765 & $= .024$ & $[-100.37, -7.21]$ & Also sig.: REM sleep, MET-med \\
PC55 & $-48.739$ & 24.523 & $= .047$ & $[-96.80, -0.67]$ & \\
PC73 & $-69.690$ & 29.171 & $= .017$ & $[-126.86, -12.52]$ & Also sig.: MET-med \\
\midrule

\multicolumn{6}{l}{\textit{Deep sleep (6 significant PCs)}} \\
PC66 & $+0.006$ & 0.002 & $= .008$ & $[+0.00, +0.01]$ & \\
PC46 & $+0.004$ & 0.002 & $= .013$ & $[+0.00, +0.01]$ & \\
PC82 & $+0.005$ & 0.002 & $= .031$ & $[+0.00, +0.01]$ & \\
PC41 & $+0.003$ & 0.002 & $= .039$ & $[+0.00, +0.01]$ & \\
PC64 & $+0.004$ & 0.002 & $= .042$ & $[+0.00, +0.01]$ & \\
PC30 & $-0.004$ & 0.001 & $= .006$ & $[-0.01, -0.00]$ & Also sig.: sleep efficiency \\
\midrule

\multicolumn{6}{l}{\textit{Sleep efficiency (10 significant PCs)}} \\
PC20 & $+0.058$ & 0.017 & $= .001$ & $[+0.03, +0.09]$ & Also sig.: onset lat., steps/day \\
PC22 & $+0.038$ & 0.017 & $= .021$ & $[+0.01, +0.07]$ & Also sig.: REM sleep \\
PC31 & $+0.056$ & 0.020 & $= .004$ & $[+0.02, +0.10]$ & Also sig.: REM sleep, RMSSD, MET-high \\
PC10 & $-0.039$ & 0.013 & $= .003$ & $[-0.07, -0.01]$ & \\
PC18 & $-0.035$ & 0.016 & $= .024$ & $[-0.07, -0.01]$ & \\
PC26 & $-0.043$ & 0.018 & $= .020$ & $[-0.08, -0.01]$ & \\
PC30 & $-0.051$ & 0.020 & $= .009$ & $[-0.09, -0.01]$ & Also sig.: deep sleep \\
PC75 & $-0.068$ & 0.031 & $= .029$ & $[-0.13, -0.01]$ & \\
PC87 & $-0.069$ & 0.034 & $= .044$ & $[-0.14, -0.00]$ & \\
PC7 & $-0.024$ & 0.012 & $= .049$ & $[-0.05, +0.00]$ & Also sig.: RMSSD \\
\midrule

\multicolumn{6}{l}{\textit{Sleep onset latency (7 significant PCs)}} \\
PC72 & $+0.172$ & 0.067 & $= .010$ & $[+0.04, +0.30]$ & \\
PC24 & $+0.092$ & 0.039 & $= .017$ & $[+0.02, +0.17]$ & \\
PC38 & $+0.111$ & 0.047 & $= .018$ & $[+0.02, +0.20]$ & \\
PC71 & $+0.140$ & 0.067 & $= .037$ & $[+0.01, +0.27]$ & Also sig.: RMSSD \\
PC91 & $-0.166$ & 0.078 & $= .032$ & $[-0.32, -0.01]$ & \\
PC54 & $-0.118$ & 0.059 & $= .044$ & $[-0.23, -0.00]$ & Also sig.: sleep duration \\
PC20 & $-0.073$ & 0.037 & $= .045$ & $[-0.14, -0.00]$ & Also sig.: sleep eff., steps/day \\
\midrule

\multicolumn{6}{l}{\textit{REM sleep (4 significant PCs)}} \\
PC22 & $+0.004$ & 0.001 & $= .008$ & $[+0.00, +0.01]$ & Also sig.: sleep efficiency \\
PC31 & $+0.003$ & 0.002 & $= .035$ & $[+0.00, +0.01]$ & Also sig.: sleep eff., RMSSD, MET-high \\
PC52 & $-0.004$ & 0.002 & $= .044$ & $[-0.01, +0.00]$ & Also sig.: steps/day, MET-med \\
PC97 & $-0.006$ & 0.003 & $= .044$ & $[-0.01, +0.00]$ & Also sig.: sleep duration \\
\midrule

\multicolumn{6}{l}{\textit{RMSSD (10 significant PCs)}} \\
PC31 & $+0.244$ & 0.084 & $= .004$ & $[+0.08, +0.41]$ & Also sig.: sleep eff., REM, MET-high \\
PC71 & $+0.322$ & 0.130 & $= .013$ & $[+0.07, +0.58]$ & Also sig.: onset latency \\
PC29 & $+0.194$ & 0.082 & $= .019$ & $[+0.03, +0.36]$ & Also sig.: steps/day, MET-high \\
PC27 & $+0.171$ & 0.080 & $= .032$ & $[+0.01, +0.33]$ & \\
PC85 & $+0.303$ & 0.147 & $= .039$ & $[+0.02, +0.59]$ & Also sig.: MET-high \\
PC60 & $+0.242$ & 0.119 & $= .042$ & $[+0.01, +0.48]$ & \\
PC7 & $-0.137$ & 0.053 & $= .010$ & $[-0.24, -0.03]$ & Also sig.: sleep efficiency \\
PC39 & $-0.240$ & 0.094 & $= .011$ & $[-0.42, -0.06]$ & \\
PC45 & $-0.209$ & 0.100 & $= .037$ & $[-0.41, -0.01]$ & \\
PC47 & $-0.204$ & 0.104 & $= .049$ & $[-0.41, -0.00]$ & Also sig.: steps/day, MET-med \\
\midrule

\multicolumn{6}{l}{\textit{Sleep duration (4 significant PCs)}} \\
PC57 & $+0.013$ & 0.006 & $= .020$ & $[+0.00, +0.03]$ & \\
PC54 & $-0.014$ & 0.006 & $= .018$ & $[-0.03, -0.00]$ & Also sig.: onset latency \\
PC97 & $-0.019$ & 0.008 & $= .019$ & $[-0.04, -0.00]$ & Also sig.: REM sleep \\
PC35 & $-0.010$ & 0.004 & $= .020$ & $[-0.02, -0.00]$ & Also sig.: steps/day \\

\end{longtable}

\noindent\textit{Note.} As in Table \ref{tab:roberta-full}, except that each joint model contains all 100 MentalRoBERTa principal components (900 tests; about 45 nominally significant results expected by chance). Abbreviations as in Table 12.

\section{Zero-Shot Domain Classification: Full Results}
\label{app:domain}

Table~\ref{tab:domain-full} reports FDR-corrected $p$-values for all 72 zero-shot domain classification tests across eight concern domains and nine wearable outcomes. One association survived FDR correction: the physical health domain was negatively associated with sleep efficiency ($\beta = -1.79$, $SE = 0.46$, $p_{\text{fdr}} = .008$). All remaining associations were non-significant after correction.

\begin{table}[H]
\centering
\caption{Zero-Shot Domain Classification: FDR-Corrected $p$-Values Across All Domain--Outcome Combinations}
\label{tab:domain-full}
\begin{tabular}{lrrrrrrrrr}
\toprule
Domain & \rotatebox{60}{Deep sleep} & \rotatebox{60}{MET high} & \rotatebox{60}{MET med.} & \rotatebox{60}{REM sleep} & \rotatebox{60}{RMSSD} & \rotatebox{60}{Sleep dur.} & \rotatebox{60}{Sleep eff.} & \rotatebox{60}{Onset lat.} & \rotatebox{60}{Steps/day} \\
\midrule
Physical health   & .073 & .564 & .839 & .564 & .564 & .669 & \textbf{.008}$^{*}$ & .824 & .714 \\
Relationships     & .564 & .839 & .576 & .564 & .971 & .564 & .565 & .971 & .613 \\
Academic workload & .564 & .579 & .839 & .640 & .839 & .564 & .564 & .839 & .839 \\
Housing           & .564 & .817 & .839 & .858 & .839 & .918 & .564 & .564 & .839 \\
General stress    & .564 & .839 & .576 & .564 & .839 & .564 & .564 & .653 & .682 \\
Future plans      & .586 & .839 & .839 & .930 & .839 & .839 & .911 & .769 & .682 \\
Money             & .640 & .576 & .971 & .672 & .918 & .576 & .564 & .768 & .971 \\
Mental health     & .817 & .669 & .973 & .930 & .839 & .839 & .817 & .768 & .839 \\
\bottomrule
\end{tabular}
\begin{tablenotes}
\small
\item \textit{Note.} Values are FDR-corrected $p$-values (Benjamini--Hochberg) across all 72 domain--outcome tests. One joint model per outcome, fitted by REML: outcome $\sim$ the eight domain probabilities (entered simultaneously; ``no concern'' omitted as the reference category because the probabilities sum to 1) + \texttt{Week\_z} + (1 $|$ \texttt{record\_id}), run on concern-present rows only ($N = 2{,}982$ sleep outcomes; $N = 2{,}966$ activity outcomes). $^{*}$Survives FDR correction at $\alpha = .05$; $\beta = -1.79$, $SE = 0.46$. Sleep eff. = sleep efficiency; Sleep dur. = sleep duration; Onset lat. = sleep onset latency; MET med. = MET-min medium; MET high = MET-min high. Domain probabilities are raw 0--1 zero-shot scores from \texttt{facebook/bart-large-mnli}, so coefficients are per full $0 \rightarrow 1$ change in probability.
\end{tablenotes}
\end{table}

\section{Additional Embedding Visualizations}
\label{app:supp-figure}
\begin{figure}[htbp]
\centering
\includegraphics[width=0.82\textwidth]{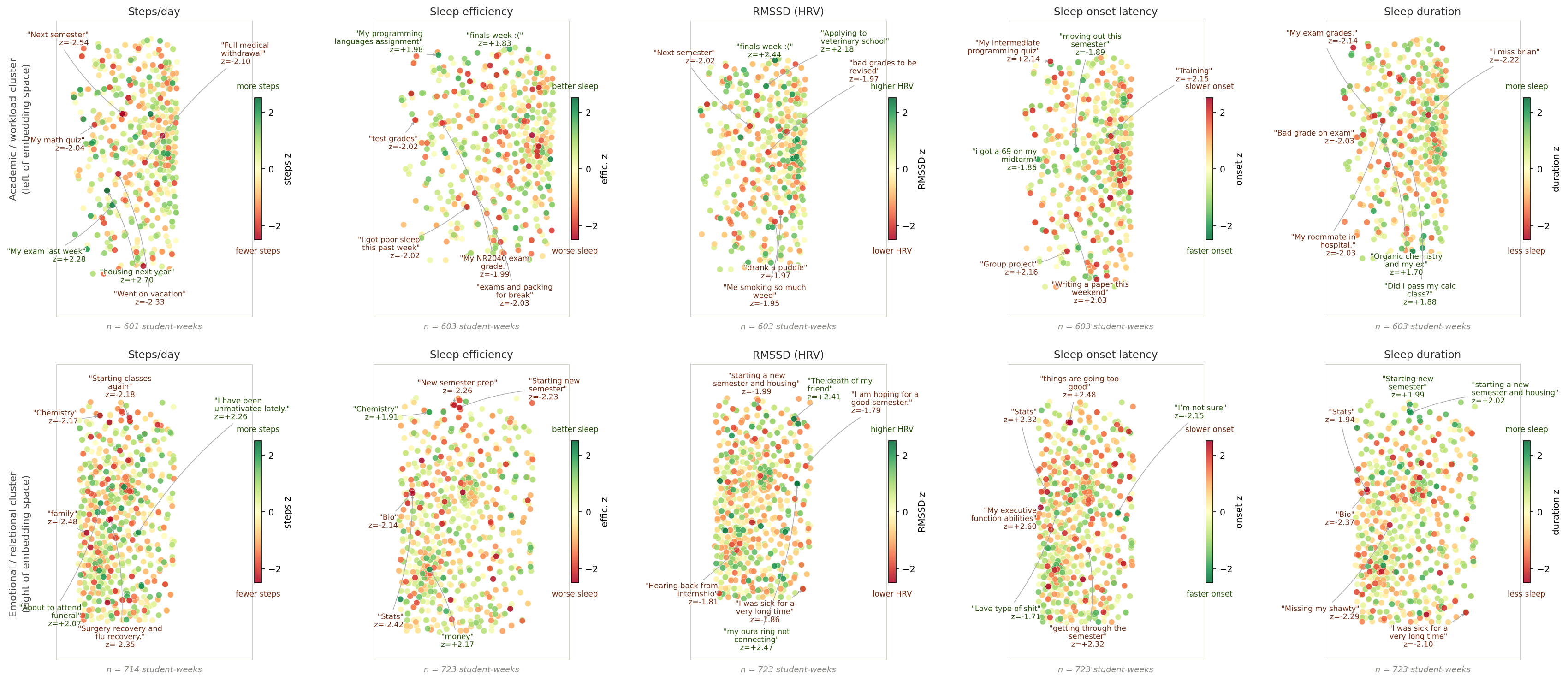}
\caption{Within-person wearable outcomes across all five outcomes in the academic/workload cluster (top row) and emotional/relational cluster (bottom row) of the RoBERTa-base embedding space. Each dot represents one student-week, colored by within-person $z$-scored outcome (green\,=\,above the student's own weekly average; red\,=\,below). Axes are unlabeled t-SNE latent dimensions. Annotated concern texts illustrate the extremes of each outcome distribution within each cluster. Academic cluster: $n = 601$--$603$ student-weeks. Emotional/relational cluster: $n = 714$--$723$ student-weeks.}
\Description{A grid of ten scatter plots arranged in two rows and five columns. Each column corresponds to one wearable outcome; the top row shows student-weeks from the academic/workload cluster (n=601-603) and the bottom row shows the emotional/relational cluster (n=714-723) of the RoBERTa-base embedding space, plotted on unlabeled t-SNE axes. In each panel, dots represent individual student-weeks colored by within-person z-scored outcome value, green for above the student's own average and red for below. A small number of points in each panel are annotated with brief concern-text excerpts marking the extremes of that outcome's distribution within the cluster.}
\label{fig:supp-all-outcomes}
\end{figure}

Figure~\ref{fig:supp-all-outcomes} extends Figure~3 in the main text by showing all five wearable outcomes across both semantic neighborhoods of the RoBERTa-base embedding space.
\FloatBarrier

\clearpage
\section{Variance Decomposition by NLP Method}
\label{app:variance-figure}
\begin{figure}[htbp]
\centering
\includegraphics[width=\textwidth]{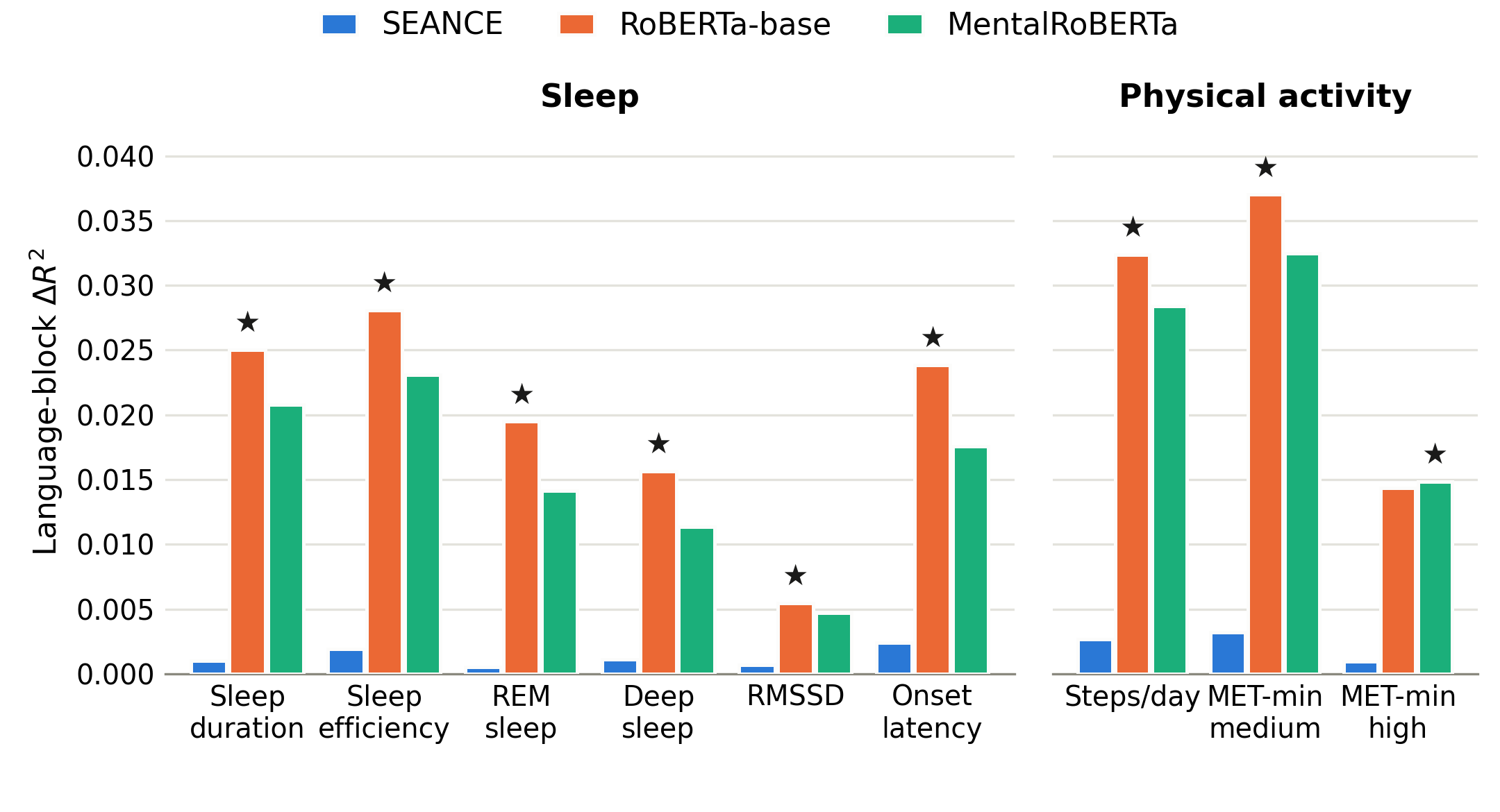}
\caption{Incremental variance explained ($\Delta R^2$) by each NLP method above semester timing across nine wearable outcomes. Bars show the language block $\Delta R^2$ for SEANCE (blue), RoBERTa-base (orange), and MentalRoBERTa (green). Stars indicate the best-performing method per outcome. All models restricted to concern-present waves ($N = 2{,}982$ sleep; $N = 2{,}966$ activity).}
\Description{Grouped bar chart with nine outcome groups on the x-axis (sleep duration, sleep efficiency, REM sleep, deep sleep, RMSSD, sleep onset latency, steps/day, MET-min medium, MET-min high), each with three bars showing language-block delta R-squared for SEANCE (blue), RoBERTa-base (orange), and MentalRoBERTa (green). Across all nine outcomes, the two embedding-based methods substantially outperform SEANCE, whose bars stay below 0.004 throughout. RoBERTa-base is the tallest or tied-tallest bar in every group, ranging from about 0.005 for RMSSD up to about 0.037 for MET-min medium; MentalRoBERTa tracks closely behind it in most groups. A star marks RoBERTa-base as the best-performing method for seven of the nine outcomes; for steps/day and MET-min high, RoBERTa-base and MentalRoBERTa are close enough to be labeled comparable rather than one clearly ahead.}
\label{fig:variance-decomp}
\end{figure}

Figure~\ref{fig:variance-decomp} provides a graphical summary of Table \ref{tab:variance} in the main text, showing the incremental variance explained ($\Delta R^2$) by each NLP method above semester timing across all nine wearable outcomes.









\end{document}